\RequirePackage{fix-cm}
\documentclass[smallextended]{svjour3}       % onecolumn (second format)
\smartqed  % flush right qed marks, e.g. at end of proof
\usepackage{graphicx}
\usepackage{balance}
\usepackage{url}
\usepackage{adjustbox}
\usepackage[utf8]{inputenc} 
\usepackage{todonotes}
\usepackage{natbib}
\usepackage[linesnumbered,ruled]{algorithm2e}
\usepackage{enumitem} 
\usepackage{nomencl}
\usepackage{subfigure}
\usepackage[flushleft]{threeparttable}
\usepackage{xcolor}
\usepackage{scalefnt}
\usepackage{adjustbox}
\usepackage{framed}
\usepackage{float}
\usepackage{xspace}
\usepackage{comment}
\usepackage{siunitx,booktabs}
\usepackage{multirow,hhline,graphicx,array}
\usepackage[most]{tcolorbox}
\usepackage[strict]{changepage}
\usepackage{color, colortbl}
\usepackage[htt]{hyphenat}
\usepackage{framed}

\definecolor{formalshade}{rgb}{0.95,0.95,1}

\newenvironment{formal}{%
  \MakeFramed{\advance\hsize-\width\FrameRestore}%
  \noindent\hspace{-4.55pt}% disable indenting first paragraph
  \begin{adjustwidth}{}{5pt}%
}
{%
  \vspace{1pt}\end{adjustwidth}\endMakeFramed%
}

\newcolumntype{L}[1]{>{\raggedright\let\newline\\\arraybackslash\hspace{0pt}}m{#1}}
\newcolumntype{C}[1]{>{\centering\let\newline\\\arraybackslash\hspace{0pt}}m{#1}}
\newcolumntype{R}[1]{>{\raggedleft\let\newline\\\arraybackslash\hspace{0pt}}m{#1}}

\newcommand{\MyPara}[1]{\vspace{.2em}\noindent\textit{\textbf{#1}}\hspace{.3em}}

\definecolor{Gray}{gray}{0.9}
\begin{document}

\title{How do annotations affect Java code readability?}

%\thanks{This work is partially supported by FAPESP (grant \#2019/12743-4), CNPq/MCTI/FNDCT (grant \#408812/2021-4), and MCTIC/CGI/FAPESP (grant \#2021/06662-1) }



% \author{Everaldo Gomes}
% \affiliation{%
%   \institution{Federal University of ABC }
%   \city{Santo Andre}
%   \country{Brazil}
% }

% \author{Jeferson Ferreira}
% \affiliation{%
%  \institution{The Federal University of Itajubá}
%  \city{Itajuba}
%  \country{Brazil}
%  }

% \author{Phyllipe Lima}
% \affiliation{%
%   \institution{The Federal University of Itajubá}
%   \city{Itajuba}
%   \country{Brazil}
%   }
% \email{phyllipe@inatel.br}

% \author{Igor Wiese}
% \affiliation{%
%   \institution{UTFPR -Federal University of Technology – Parana}
%   \city{Campo Mourao}
%   \country{Brazil}
%   }
% \email{igor@utfpr.edu.br}

% \author{Marco Gerosa}
% \affiliation{%
%   \institution{NAU - Northern Arizona University}
%   \city{Flagstaff}
% \email{marco.gerosa@nau.edu}

% \author{Paulo Meirelles}
% \affiliation{%
%   \institution{Federal University of ABC}
%   \city{Santo Andre}
%   \country{Brazil}}
% \email{paulo.meirelles@ufabc.edu.br}

\author{Eduardo Guerra \and
        Everaldo Gomes \and 
        Jeferson Ferreira \and
        Igor Wiese \and
        Phyllipe Lima \and
        Marco Gerosa \and
        Paulo Meirelles.
}


\institute{ E. Guerra. 
              Free University of Bozen-Bolzano, Italy.
              \email{eduardo.guerra@unibz.it}
            \\
            E. Gomes.
                University of São Paulo, Brazil.
            \email{everaldogjr@usp.br}
            \\
            J. Ferreira.
                SONDA, Brazil.
                \email{jeferson.ferreira@sonda.com}
            \\
            I. Wiese.
              Federal Technological University of Paran\'a, Brazil.
                \email{igor@utfpr.edu.br}
            \\
            P. Lima.
              Federal University of Itajub\'a, Brazil.
                \email{phyllipe@unifei.edu.br}
            \\
           M. Gerosa.
              Northern Arizona University, USA.
                \email{marco.gerosa@nau.edu}
            \\
           P. Meirelles.
              University of São Paulo, Brazil.
                \email{paulormm@ime.usp.br}
             \\
          }

\date{Received: date / Accepted: date}

\maketitle

\begin{abstract}
\parindent=0mm \color{white}.\color{black}

\textbf{Context:}
Code annotations have gained widespread popularity in programming languages, offering developers the ability to attach metadata to code elements to define custom behaviors. Many modern frameworks and APIs use annotations to keep integration less verbose and located nearer to the corresponding code element. Despite these advantages, practitioners' anecdotal evidence suggests that annotations might negatively affect code readability.
\textbf{Objective:}
To better understand this effect, this paper systematically investigates the relationship between code annotations and code readability.
\textbf{Method:}
In a survey with software developers (n=332), we present 15 pairs of Java code snippets with and without code annotations. These pairs were designed considering five categories of annotation used in real-world Java frameworks and APIs. Survey participants selected the code snippet they considered more readable for each pair and answered an open question about how annotations affect the code's readability.
\textbf{Results:}
Preferences were scattered for all categories of annotation usage, revealing no consensus among participants. The answers were spread even when segregated by participants' programming or annotation-related experience. Nevertheless, some participants showed a consistent preference in favor or against annotations across all categories, which may indicate a personal preference. Our qualitative analysis of the open-ended questions revealed that participants often praise annotation impacts on design, maintainability, and productivity but expressed contrasting views on understandability and code clarity.
\textbf{Conclusions:}
Software developers and API designers can consider our results when deciding whether to use annotations, equipped with the insight that developers express contrasting views of the annotations' impact on code readability. 

\keywords{Code Annotations \and Program Comprehension \and Code Readability \and Software Engineering}
%\subclass{MSC code1 \and MSC code2 \and more}
\end{abstract}

\section{Introduction}

%just annotations
Code annotation is a programming language feature that allows adding custom metadata directly in the source code. The Java language introduced annotations in 2004 \citep{JSR2004} and, since then, several popular APIs and frameworks have used this approach, resulting in a significant number of projects with annotations \citep{lima2018jss, Yu2019}. Annotation enables declarative statements in source code that otherwise would be defined programmatically or by configuration files \citep{Fernandes2010}. While annotations potentially benefit maintenance by allowing metadata to be defined proximate to the relevant code element, they also add new information and identifiers to the source code, which can contribute to developers' cognitive overload \citep{Wulff2019} and harm readability.

%just readability
Code readability is essential to program comprehension and maintainability \citep{lucas2019does, Rugaber00,Buse2010}. Readability is especially important for newcomers to a project since they still do not have the contextual knowledge that helps to understand the source code \citep{steinmacher2015social}. Previous studies present evidence that the presence of certain programming features might have a positive or negative impact on readability \citep{tashtoush2013impact,dos2018impacts}. Consequently, knowing how particular features such as code annotations impact readability might guide API designers in choosing which features to adopt.
% , since factors like complexity, usage of design concepts, formatting, and source code lexicon have been widely recognized as elements that impact program understanding

%and is considered a subjective code property \citep{Posnett2011}

%other studies
Recent studies revealed that some projects abuse annotations \citep{lima2018jss} and that a high number of identifiers in the code harms readability \citep{Buse2010, tashtoush2013impact}. On the positive side, an analysis of the code evolution \citep{Yu2019} pointed out that annotated Java code tends to be less error-prone. Given the relationship between readability and code maintenance, it is reasonable to consider whether annotations might have that effect due to their impact on code readability. This idea has been discussed by developers in blog posts \citep{StackOverflow-AgainstAnnotations,Bugayenko2016,Warski2017}, with many claiming that annotations can potentially harm source code readability. 

Nevertheless, no previous scientific work has investigated the effects of code annotations on readability. Without such a study, it is unclear whether these statements posted in the gray literature represent a general perception or are just individual opinions supported by several potentially biased comments posted by other developers. The lack of evidence in the scientific literature about the impact of annotations on readability, and more specifically, how this feature affects particular audiences and usage scenarios, might lead designers to make decisions based on personal perceptions. 

%annotations + readability (importance)
%Evidence that annotations receive several changes in software evolution \citep{Yu2019} highlight the importance of their impact on code readability. 

Our study aims to fill this gap by investigating the impact of annotations on code readability from the perspective of various software developers. We designed a survey considering five different annotation usage categories, defined from documented patterns \citep{guerra2010architectural,guerra2016design,yang2008empirical}. To ensure the relevancy of these categories, we evaluated their occurrence in existing frameworks and APIs by mining software repositories and manually classifying the annotations. We found a high number of occurrences for each category and used them to design a survey that presents pairs of code snippets with the same behavior to the participant: one that uses code annotations and the other that uses object-oriented techniques. We also asked open questions in which participants could express their thoughts about the annotation's impact on readability and provide additional input to the research.

%We evaluated the categories via a study that extracted annotation schemas from open-source projects and classified the annotations of 42 frameworks and APIs. As a result, we found that all annotations could be classified into one of the five categories and that all received a significant number of occurrences. We considered all five categories in the survey design based on these results.

Our survey received 332 valid answers from developers with different profiles, considering programming experience, Java expertise, and annotation familiarity. Surprisingly, the answers to all questions followed a normal distribution without a clear positive or negative trend. Our analysis did not reveal a correlation between the responses, participant profiles, or annotation usage categories. However, we found that around half of the participants have a consistent preference in favor or against annotations, which might explain some posts in the gray literature discouraging the usage of annotations and a large number of projects adopting this feature. Our respondents provided a number of recommendations, including avoiding overusing annotation, keeping the annotation usage simple, and choosing good names.

The main contribution of this paper is investigating how annotations affect readability, considering different developer profiles and annotation usages. Several new standard APIs and frameworks adopt an annotation-based API instead of an object-oriented one, which is an important design choice. Designers and users of APIs can use our results to support their design decisions when choosing between an annotation-based or a classic object-oriented API and spending extra effort on improving readability given that there are conflicting preferences. As a secondary contribution, this work evaluated the frequency of a set of annotation usage categories in a sample of APIs used by open-source projects.

\section{Code Annotations}

Code annotations are a type of metadata in the source code. The term metadata is used in various contexts in computer science, meaning data about the data itself. In the object-oriented context, metadata includes the information that describes a given class. Some tools or frameworks can consume metadata and execute routines based on class structure. For instance, metadata can be used for source code generation \citep{Damyanov2004}, compile-time verification \citep{Ernst2008,Quinonez2008}, class transformation\footnote{\url{http://projectlombok.org}}, and framework adaptation \citep{Guerra2010b}. 

Metadata can be defined through the usage of object-oriented techniques. For example, the presence of an interface can define a class property---this practice is called marking interface \citep{bloch2016effective}. Class metadata can also be returned as fixed information from static methods or variables with name conventions. Alternatively, an object containing the metadata might be created programmatically by setting the appropriate information \citep{guerra2013reference}. Finally, as another option, metadata can be passed using fixed literals as arguments to methods.

External resources, such as configuration files or databases, can be used to define custom metadata \citep{Fernandes2010}. The drawback of this approach is the distance between the metadata and the referred code element. This approach also adds some verbosity, since a complete path to the element must be provided for the framework to correctly consume the metadata. Some frameworks, like Ruby on Rails and CakePHP, use an alternative to define behavior through code conventions \citep{Chen}. Although this choice can be productive in some contexts, code conventions have limited expressiveness. Moreover, since the metadata is hidden behind the conventions, an unwary developer might alter it, producing unwanted effects.

Some programming languages provide features that allow custom metadata to be defined and included directly on programming elements through code annotations. This feature is supported in languages such as Java, through annotations, and in C\#, through attributes. Since code annotations are located nearer to the programming element, their use can be less verbose than an external definition, since the context is already well established. 

\subsection{Code Annotations in Java} \label{sec:annotationsinjava}

\textbf{Annotations} were introduced as a language feature in Java 1.5. Some standard APIs like EJB 3.0 (Enterprise Java Beans), JPA (Java Persistence API), and CDI (Context and Dependency Injection) extensively use metadata in the form of annotations. This native annotation support encouraged Java frameworks and API developers to adopt the metadata-based approach in their solutions, confirming the tendency to keep the metadata files inside the source code instead of using separate files \citep{CORDOBASANCHEZ2016}.

An annotation-based API, or metadata-based framework, defines and exposes a set of annotations to application developers to configure programming elements and execute the desired behavior. This set of annotations defining a given domain's metadata structure for an API is called \textbf{annotation schema} \citep{lima2018jss}. For instance, the annotation schema for object-relational mapping of the JPA API includes annotations like \texttt{@Table}, \texttt{@Column}, \texttt{@Id}, and \texttt{@OneToMany}.

The code annotations from the same schema are usually located in the same Java package. In the example of the JPA API, all the object-relational mapping annotations are located on the \texttt{javax.persistence} package. Therefore, \cite{lima2018jss} suggests the heuristic of using the package to identify and name the annotation schema. Accordingly, we can say that \texttt{@Table} and \texttt{@Column} belong to the annotation schema \texttt{javax.persistence}. Moreover, the same API or framework might define multiple packages with annotations representing metadata with different purposes. In this case, each package represents a different annotation schema. For instance, the \texttt{javax.persistence.metamodel} is an example of another annotation schema that belongs to the JPA API. In this paper, we adopt the same heuristic proposed by \cite{lima2018jss} and use the annotation package to identify its annotation schema.

\subsection{The Impact of Code Annotations on Program Comprehension}
\label{sec:readability}

Program Comprehension is essential to software maintenance and is directly linked to adequate software evolution \citep{STOREY2000183}. Comprehending the source code and the structure of a system is required before applying any proper modification or enhancement \citep{7961527}. Developers can use different artifacts to help to comprehend a program, including the source code \citep{STOREY2000183}, which should offer high readability. Therefore, code readability is an essential component of program comprehension \citep{lucas2019does, Rugaber00}.

Readability can be defined as \textit{``a human judgment of how easy a text is to understand''} \citep{Buse2010} and is directly related to its maintainability \citep{Buse2010}. Development teams pursue this quality attribute since the typical software product life-cycle cost distribution is 70\% maintenance and 30\% development \citep{Boehm2001}. The impact of readability on maintenance is high since reading source code is the most time-consuming process of all maintenance activities \citep{Raymond1991, Rugaber00}. On the one hand, metadata (such as the one offered by code annotation) can enhance the readability of the source code by adding relevant information for its comprehension. On the other hand, adding metadata with new identifiers increases developers’ cognitive load  \citep{Wulff2019} and might have a negative impact.

Readability is considered a subjective code property \citep{Posnett2011}. \cite{scalabrino2017automatically} showed that objective metrics that assess code readability do not correlate with its understandability, and \cite{pantiuchina2018improving} showed that such metrics do not capture code quality differences as perceived by developers. Therefore, many studies, like ours, investigate code readability through surveys in which participants compare the readability of pairs of code snippets. For example, \cite{dos2018impacts} conducted a survey that assessed the impact of Java coding practices and conventions on readability by showing pairs of code snippets to software developers. Their results helped to identify practices and conventions with positive and negative impacts on readability, but they did not investigate code annotations. \cite{lucas2019does} evaluated the effects of lambda expressions on Java code comprehension. Their study employed an online survey in which participants evaluated pairs of code snippets. The survey indicated that introducing lambda expressions on legacy code improves its readability. Therefore, the literature supplies plenty of evidence that certain programming features might positively or negatively impact code readability \citep{tashtoush2013impact}. 

%%%%% related work
\subsection{Impacts of code annotation on other quality attributes}

Although no previous work directly investigates the impacts of code annotation on code readability, there are studies that investigated tangential aspects, such as maintainability \citep{Buse2010} and the number of identifiers present in the code \citep{tashtoush2013impact}. 

\cite{guerra2013qualitative} assessed the impact of frameworks based on annotations compared to object-oriented frameworks. Similar to the present work, Guerra and Fernandes considered categories of annotation usage to design scenarios. Their study did not find evidence that annotations lead to reduced coupling and that the indirection of metadata definition leads to difficulty in debugging.  

\cite{lima2018jss} proposed a suite of software metrics to characterize code annotations' complexity, coupling, and size. Based on data from 25 open-source Java projects, they observed that 78\% of classes contain at least one annotation, highlighting the relevance of studying the impact of their presence. Identifying metric outliers, such as a class with 729 code annotations and one annotation that took 58 lines of code, reveals abuses in using this language feature. 

\cite{Yu2019} performed a large-scale empirical study on code annotations usage, evolution, and impact. The authors collected data from 1,094 open-source Java projects and conducted a historical analysis to assess code annotations. The study revealed many changes in annotations during project evolution, implicating that some developers tend to use annotations subjectively and arbitrarily, introducing code problems. The study also revealed that developers intending to improve code readability add annotations to the existing program elements. By relating annotation usage to code error-proneness, the study concluded that they could potentially enhance software quality by relating annotation usage to code error-proneness.

Concerning annotation repetition, \cite{teixeira2018does} investigated the source code of a web application, searching for annotations repeated throughout the source code. The findings revealed that some annotations were repeated around 100 times in code elements with shared characteristics in the target system. The study suggested that more general definitions, such as application-specific code conventions, could significantly reduce the number of configurations.

A recent experiment about framework development compared an annotation-based API to an object orientation-based one for metadata reading \citep{guerra2020}. The results showed a more consistent behavior in the evolution of coupling and complexity metrics with the annotation-based approach, but no significant differences in productivity. 

The work of \cite{Lima2022cadv} proposed a software visualization approach to observe how code annotations are distributed in a given system and improve code comprehension of annotation-based systems. The authors conducted an empirical evaluation with students and professional developers using a Java web application as the target system. From their findings, the authors observed a strong relationship between the presence of code annotation and the responsibility of the package/class using that annotation. This finding suggests that code annotations may also guide the architectural design of software systems.

\subsection{Code Annotations Usage Scenarios}
\label{sec:categories}

Code annotations can be used for multiple purposes. To cover a variety of scenarios in the survey, we searched the literature for patterns related to code annotations. We identified five annotation usage categories. The following subsections describe the process we followed and the usage categories. 

\subsubsection{Category Identification Process}

We surveyed pattern collections and pattern languages related to code annotations to identify the categories. The following set of patterns were considered:

\begin{itemize}

    \item \textit{A Pattern Language for Metadata-based Frameworks} \citep{Guerra2010b}: These patterns address solutions used in metadata-based frameworks.
    
    \item \textit{Architectural patterns for metadata-based frameworks usage} \citep{guerra2010architectural}: These patterns are related to the metadata usage in existing frameworks.
    
    \item \textit{Idioms for Code Annotations in the Java Language} \citep{guerra2010idioms}: These language-specific patterns document common practices in the annotation definition in the Java language.
    
    \item \textit{Design patterns for annotation-based APIs}\citep{guerra2016design}: These patterns describe recurrent solutions in the usage of code annotations for the creation of APIs.

\end{itemize}

As the inclusion criteria, we considered patterns that use code annotations as part of the solution to the proposed problem. In other words, when the solution suggests the usage of annotations to more general problems. We excluded patterns in which the usage of annotations is already part of their context and provided solutions to deal with annotations independent of the scenario. For instance, we excluded patterns that offer solutions to the internal structure of frameworks that process annotations and represent more complex metadata in the form of annotations. The inclusion and exclusion involved a discussion followed by a consensus among three authors of this paper.

Our analyses disregarded the patterns from \citep{Guerra2010b} and \citep{guerra2010idioms} since none describes usage scenarios for code annotations. The pattern language from metadata-based frameworks focuses on the internal structure of frameworks that consume annotations and can be applied to any framework independent of its domain. The idioms for annotations in Java focus on strategies to represent metadata using annotations, i.e., on the structure of the annotations. For example, some patterns document solutions using annotations to represent lists of composite objects and more complex expressions. 
%All the patterns are also domain-agnostic and can be applied to different usage scenarios.

In analyzing the architectural patterns for metadata-based framework usage \citep{guerra2010architectural}, the patterns matched our inclusion criteria, and three of the four documented patterns were included. The pattern \texttt{Metadata-based Graphical Component} was excluded because it is a specialization of one of the other patterns. In the pattern collection for annotation-based APIs \citep{guerra2016design}, three patterns are referred to using annotations in the scenarios documented in the architectural patterns. These patterns focus on the same scenario but from a different perspective. Additionally, two other patterns, \texttt{Class Stamp} and \texttt{Metadata Parametrization}, present solutions that were classified together as a fourth usage scenario.

As a preliminary evaluation, the four identified usage scenarios were used to classify the annotations of two widely used annotation APIs: JPA for object-relational mapping and Spring framework annotations for web development. For these two frameworks, all the annotations that did not fit in one of the initial four categories were used to implement the pattern \texttt{Dependency Injection} \citep{yang2008empirical}. In this pattern, a class called Injector is responsible for creating and injecting the dependency. In an implementation based on code annotations, the Injector uses them to locate the fields that should receive the injection as parameters to create the dependencies. Based on that, a fifth category was added.

The identified categories are described in the next section. To further evaluate the completeness and relevance of the categories, we performed an additional study, presented in Section \ref{sec:annotations_usage}, to evaluate how frequently annotations from each category are present in existing APIs and if we can find annotations that do not fit in any category.

\subsubsection{Identified Categories}

The following are the categories identified based on the process described in the previous section, used to guide the development of the survey. 

\textbf{Callback Method} is a category based on the patterns \texttt{Configured Method Handler} \citep{guerra2010architectural} and \texttt{Callback Configuration} \citep{guerra2016design}. Annotations in this category configure a method invoked in response to a situation or event. Annotations can contain additional information defining constraints about when it should be called. An example of this category is the annotation \texttt{@PrePersist} from the JPA API, which configures methods that should be called before an entity is persisted in the database.
    
\textbf{Information Mapping} is a category based on the patterns \texttt{Entity Mapping} \citep{guerra2010architectural} and \texttt{Method Parameter Mapping} \citep{guerra2016design}. Annotations in this category map two different representations of a system entity. Typical usages include mapping domain classes to databases or external file formats or methods to external services. Examples of this category include the annotations \texttt{@XmlElement} and \texttt{@XmlAttribute} from JAXB API, which map fields and properties from a class, respectively, to elements and attributes in an XML representation. 

\textbf{Dependency Injection} is a category based on the more general pattern \texttt{Dependency Injection} \citep{yang2008empirical}, which has an annotation-based implementation in several APIs and frameworks. This category includes annotations configuring how instances should be injected into an instantiated class. For example, the annotation \texttt{@Autowired} is used in the Spring Framework to configure fields that should receive a dependency injection. 
% For example, the annotation in a field can indicate that a specific instance should be set there.

\textbf{Proxy Configuration} is a category based on the patterns \texttt{Crosscutting Metadata Configuration} \citep{guerra2010architectural} and \texttt{Proxy Processing Configuration} \citep{guerra2016design}. Annotations in this category define constraints on processing that happens before or after a method invocation. These annotations are used by dynamic proxies, aspects \citep{guerra2008using}, or interceptor components \citep{guerra2013flexible} to parameterize their processing for a specific class or method. For example, \texttt{@TransactionAttribute} from the EJB API configures transaction propagation. Another example in the same API is \texttt{@RolesAllowed}, used to configure access control constraints.
    
\textbf{Rule Definition} is a category based on the patterns \texttt{Class Stamp} and \texttt{Metadata Parametrization} \citep{guerra2016design}. This category is related to annotations used to configure parameters for processing related to the target class. This category can be considered a more broad category since it includes any annotation that provides information for a component or a framework to execute a logic related to the annotated class. In the Bean Validation API, for example, annotations that define constraints for object validation, such as \texttt{@Size} and \texttt{@NotNull}, are part of this category.

\section{Evaluation of the Annotation Usage Scenarios}
\label{sec:annotations_usage}

As described in Section \ref{sec:categories}, code annotations can be used for multiple purposes \citep{guerra2010architectural,guerra2016design,yang2008empirical}. In this study, we aimed to conduct a comprehensive investigation that encompassed multiple real-world usage scenarios. Therefore, before designing the survey, we investigated whether the scenarios defined in Section \ref{sec:categories} cover adequately real-world usages, without including overly restrictive or irrelevant scenarios. We analyzed a sample of real-world annotation schemas to assess whether the defined categories were prevalent and comprehensive enough to be included in the study.  
%The set of categories was defined by reviewing relevant literature and organizing patterns and concepts currently dispersed in multiple works into a coherent set of categories. 

With this goal in mind, we verified the frequency with which annotations from each category are present in annotation schemas from existing APIs and frameworks. We also looked for annotations that do not fit into any category. We do not investigate the frequency with which the annotations from each category are used in classes, but we calculate the frequency at which APIs include at least one annotation of a given category in the context of our sample. Considering different approaches for taxonomy evaluation \citep{usman2017taxonomies}, this evaluation study can be classified as utility demonstration, in which its utility is demonstrated by actually classifying subject matter examples \citep{vsmite2014empirically, wheaton1968development}. 

We extracted a sample of annotation schemas used in real-world frameworks and APIs and then manually analyzed the annotations from each schema. Our goal was to evaluate at least one hundred real-world annotations. 
The annotation extraction procedure followed objective and well-defined criteria focusing on increasing the reproducibility of this study. In the following, we detail the annotations extraction and classification processes.
 
\MyPara{Annotation extraction.} To find annotation schemas, we first selected a few popular Java projects hosted on GitHub to extract the annotations used in their frameworks and APIs. We considered projects with more than 10,000 stars, with Java as the declared language, and containing files with the \texttt{.java} extension. We ranked the projects by the number of stars and individually evaluated them according to the following inclusion criteria: (a) has a license; (b) the README and the commit messages are primarily written in English; (c) is a software project (not a didactic material, for instance); (d) has at least one release; (e) the tool used to extract data from the annotations, ASniffer \citep{annotationsniffer2020}, can process its source code without errors. The first 30 projects that passed the criteria were selected for the study.

Then, we used the open-source tool Annotation Sniffer (ASniffer) \citep{annotationsniffer2020} to extract the annotation schemas and obtain code annotation characteristics from the source code. We considered the annotation package to identify an annotation schema (see Section \ref{sec:annotationsinjava}). Schemas from standard Java APIs (beginning with \texttt{Java.*}, \texttt{javax.*}, and \texttt{javafx.*}) were separated from schemas defined by third-party frameworks to allow segmented analyses. Additionally, we excluded schemas that (a) use a standard API package name but are not official, (b) with only compile-time processing annotations (out of scope), and (c) schemas for which the JavaDoc documentation could not be found on the web. 

After running the ASniffer on the selected 30 projects, we got a raw list of annotation schemas. To separate and organize them, we ran an additional script named \textit{schema-organizer}\footnote{\url{https://github.com/metaisbeta/schema-organizer}}. After running this script, we found 26 annotation schemas from standard Java APIs and 326 from third-party libraries and frameworks. Applying the exclusion criteria on the standard Java APIs, we excluded five schemas and stayed with 21 schemas.
The standard Java APIs have a well-known design process that counts with the participation of several specialists and experienced developers and have high visibility. Because of that, we consider that these APIs usually follow good design practices, which makes them suitable subjects for this study.

To balance the sample with annotation schemas from third-party libraries, we added the 21 most frequent ones that did not match any exclusion criteria, finishing the same amount of schemas from both types. Since we do not aim to evaluate precisely the frequency of each category in this population, we decided to represent Java standard APIs and third-party libraries with the same amount of annotation schemas, instead of following their percentage in the sample. Table \ref{tab:schemas_excluded} lists all the excluded annotation schemas with the respective reason. Therefore, we had a final list of 42 annotation schemas. The 21 schemas from Java APIs contained 95 annotations, and the 21 schemas from third-party libraries and frameworks had 48, resulting in 143 total annotations for analysis.

\begin{table}[t]
\caption{Excluded annotation schemas.}
\centering
\begin{tabular}{|l|l|l|}
        \hline
        \textbf{Annotation Schema} & \textbf{API Type} & \textbf{Reason} \\ \hline
        java.lang & Standard & Compile-time \\  
        java.lang.annotation & Standard & Compile-time \\
        javax.cache.annotation & Standard & Not official API\\
        javax.annotation.concurrent & Standard & Not official API\\
        javax.annotation.meta & Standard & Not official API\\
        \hline
        com.google.common.annotations &  Third-party & Compile-time \\ 
        org.jkiss.code &  Third-party & Javadoc not found\\ 
        org.openjdk.jmh.annotations &  Third-party & Compile-time \\ 
        edu.umd.cs.findbugs.annotations &  Third-party & Compile-time \\ 
        com.google.errorprone.annotations &  Third-party & Compile-time \\ 
        org.checkerframework.checker.nullness.qual &  Third-party & Compile-time \\ 
        org.jkiss.dbeaver.model.meta &  Third-party & Javadoc not found\\ 
        org.checkerframework.checker.nullness.compatqual &  Third-party & Compile-time \\ 
        io.netty.util.internal &  Third-party & Javadoc not found\\ 
        org.kohsuke.accmod &  Third-party & Javadoc not found\\ 
        org.kohsuke.stapler &  Third-party & Javadoc not found\\ 
        \hline
    \end{tabular}
    \label{tab:schemas_excluded}
\end{table}

\MyPara{Annotation classification.} Two researchers manually analyzed the annotation schemas and classified them using the five aforementioned categories as a seed. During this procedure, the researchers were attentive to annotations that did not fit into any category. The researchers attributed a category to a schema when at least one annotation fit in the category. The JavaDoc of the annotation schema package was used to identify all the annotations present in the schema. To classify each annotation, the researchers also considered its description and source code examples found on the website Tabnine \footnote{\url{https://www.tabnine.com/code}}. The researchers worked independently on all schemas and met to discuss and reach a consensus on the discrepancies. The classifications did not match for 19 annotations, representing a disagreement rate of 13\% with a Cohen's kappa score of 0.81, which can be considered an excellent agreement \citep{mchugh2012interrater}. 

%Discussing the differences made a consensus in all cases' final classification. 
%We calculated Cohen’s Kappa Statistic. This method measures the level of agreement between two raters or judges who each classify items into mutually exclusive categories. Cohen’s Kappa ranges between 0 and 1, with 0 indicating no agreement between the two raters and 1 indicating perfect agreement between the two raters \citep{mchugh2012interrater}.

This type of manual classification and labeling is frequently used in the field of software engineering, e.g., \citep{GUNAWARDENA2023107054, piantadosi2020does}. To mitigate common threats present when this approach is used, such as confirmation and researcher bias, our study adopted guidelines well accepted by the software engineering research community. For transparency, we declare that one of the researchers who worked on the classification is the author of some patterns considered for defining the categories. The other researcher who participated in the classification had no previous contact with the categories or patterns before this study. However, this researcher had a consolidated experience in Java programming and software development in general. This researcher received an initial explanation about the categories, and both researchers used as support material the categories' descriptions and the documented patterns in the literature that were used for categories definition. To minimize the effect of researcher bias, the researchers conducted the classification independently, and the differences in this initial classification were considered for calculating the agreement. After some discussion, a consensus was obtained for all annotations, and the primary sources of disagreements were documented. The researcher who worked on the annotation extraction did not participate in the classification, helping to mitigate bias. 

\subsection{Results of the categories evaluation}
\label{subsec:categories-study}

Table \ref{tab:schemas_category} presents the number of schemas containing at least one annotation of the given category. Each line represents a category, and the entries represent the respective quantity of schemas with at least one annotation that fits the category. The percentage is calculated based on the total number of schemas of each type, which is 21 for standard APIs, 21 for third-party frameworks, and 42 in total. As can be noticed, all annotations were classified into five categories and none was marked as ``Other.''

\begin{table}[t]
\caption{Number of schemas with at least one annotation pertaining to the category. The analysis was segmented into standard Java APIs and third-party frameworks. The total does not sum 100\% since a schema can contain annotations from more than one category.}
\centering
\begin{tabular}{|l|r|r|r|}
        \hline
        \textbf{Category} & \textbf{Java APIs} & \textbf{Frameworks} & \textbf{Total} \\ \hline
        Callback Method & 4 (19\%) & 4 (19\%) & 8 (19\%)\\  
        Information Mapping & 10 (48\%) & 6 (29\%) & 16 (38\%)\\ 
        Dependency Injection & 6 (29\%) & 5 (24\%) & 11 (26\%)\\ 
        Proxy Configuration & 7 (33\%) & 3 (14\%) & 10 (24\%)\\ 
        Rule Definition & 17 (81\%) & 19 (90\%) & 36 (86\%)\\
        Other & 0 (0\%) & 0 (0\%) & 0 (0\%)\\
        \hline
    \end{tabular}
    \label{tab:schemas_category}
\end{table}

%\subsubsection{Discussion}

Based on this analysis, we conclude that the proposed categories are comprehensive enough to classify commonly used annotations, i.e., all the annotations included in our sample could be classified into one of the categories. We also conclude that all the categories have a significant enough number of occurrences to be included in our study---the minimum number of occurrences was 19\% for the \texttt{Callback Method} category. We expected to discard categories with less than 5\% of occurrences, which did not occur.  

We highlight that this study did not aim to determine the exact frequency of each category in annotation schemas or to compare the frequency of annotations from each category in standard Java APIs and frameworks. The percentages reported in Table \ref{tab:schemas_category} apply to the analyzed sample and should not be generalized.

%We could not find a runtime-consumed annotation that does not belong to one of the five categories, and no category had a negligible number of occurrences (the minimum was 19\%). 
%Consequently, future proposals for new categories should probably be subdivisions of the existing ones. A new category that does not fit in one of these would not have significant usage in existing frameworks and APIs.

%Based on the data presented in Table \ref{tab:schemas_category}, we consider that all categories had significant occurrences in the evaluated annotation schemas. The category with fewer occurrences was the Callback Method, with eight schemas (19\%) with at least one annotation of this category.  

The category \textit{Rule Definition} had many more occurrences than the others. We observed that it is common to have this kind of annotation to configure general parameters, even when the primary goal of the whole annotation schema fits in one of the other categories. Future studies might investigate the occurrences of annotations classified in this category and try to identify more specific usages and define subdivisions of it.

There were some cases in which the classification of the annotation usage was ambiguous. For example, \textit{Information Mapping} and \textit{Callback Method} categories have some intersection when the mapping occurs between methods and services from other APIs. The uncertainty is that the method is called due to an event (\textit{Callback Method}) but mapped from a request received from another API. In other instances, \textit{Information Mapping} seems to overlap Dependency Injection. This is especially true when what is being mapped is also injected into the target class. The researchers discussed and defined the classification in those cases by considering the overall context. While classifying each annotation in a single category was suitable for this study's goals, future studies can explore this overlap and interaction between the categories.

%Some insights presented in this discussion about the categories are unrelated to the readability study. However, we choose to include these secondary results since sources of disagreement are relevant to analyze the research bias threat from the survey study, and they can bring relevant insights for future studies. 

\section{Survey design}
\label{sec:research}

Code annotations define behavior and logic differently, using a declarative approach for metadata definition. This metadata can replace some code previously defined imperatively or inside object-oriented definitions. This paradigm change might affect code readability in a way that the current metrics used for code readability cannot capture. Before delving into the metrics influencing code readability, it is essential to acknowledge that specific conventional metrics fall short in capturing characteristics introduced by code annotations. For example, traditional complexity and coupling metrics inadequately represent the complexity and coupling introduced by annotations~\citep{guerra2009questioning}. The readability metrics~\citep{Buse2010, Posnett2011, tashtoush2013impact} mostly rely on features related to vocabulary, identifiers, code size, and control flow. None of these features captures the declarative aspect of the annotation approach that might affect code readability since it is possible to have codes with the same behavior, which uses and does not use annotations, with similar size, similar number of identifiers, same control flow and that share the same vocabulary.

The primary goal of our study is to assess the impact of annotations on code readability. Aligned with the vision that readability is better measured subjectively~\citep{Posnett2011, scalabrino2017automatically, pantiuchina2018improving} and considering that the current readability metrics do not capture some characteristics of annotations, we surveyed software developers to collect their perspectives and preferences.

The survey design was inspired by previous work \citep{dos2018impacts,lucas2019does}, which offered pairs of code snippets to participants. Each pair contained an annotated code and an object-oriented alternative that produces the same behavior. The survey participants selected the snippet that they considered more readable. Since our intended audience includes developers with and without code annotation experience, we posit that comparing code snippets is the most suitable way to design this survey since participants would know an alternative implementation for the same behavior. A developer without more advanced knowledge would have difficulty seeing an alternative to the code annotations, which would interfere with the readability assessment. Moreover, we determined that this approach reduces potential threats (e.g., maturation, memory), allowing the participants to see both alternatives simultaneously and choose one. 

%Our decision to provide pairs of code snippets to be compared instead of providing only one to be classified is also justified by our intent to include participants with and without familiarity with annotations. While most Java developers already configured annotations from an existing API, few know how to create and consume them. Indeed, it is not common to have that topic in introductory programming courses and books. Moreover, the usage of the Reflection API, which is considered an advanced programming topic in Java, is required to read annotations from code elements. Consequently, a developer without this more advanced knowledge would have difficulty seeing an alternative to the code annotations, which would interfere with the readability assessment. In other words, it is hard to judge the impact of using a programming feature if another way to achieve the same result is unknown. 

As presented in Section \ref{sec:categories}, code annotations are used in multiple scenarios. Our study investigated whether the usage scenarios influence the annotations' impact on code readability. We included all relevant annotation usage scenarios, as defined in Section \ref{sec:categories}. These categories focus on annotations consumed at runtime by frameworks and APIs. Code annotations can be consumed directly from the class source code, from the generated bytecode after compile time or at load time, or at runtime using reflection. This work focuses on the last case, when annotations are consumed at runtime as a metadata source for frameworks and APIs. The choice to focus on runtime annotations is justified by the reference used for comparison, since annotations consumed by compile-time that perform verifications \citep{dietl2011building} might not have an apparent equivalent solution using only object-oriented alternatives. Moreover, the focus is on the regular usage of annotations and not on scenarios that abuse this programming feature, as sometimes occurs in practice \citep{lima2018jss}. 

%Therefore, we included questions representing annotations from the defined categories, which cover the practical usage of the annotations. 

To collect a more qualitative perspective, we also asked the participants open questions wherein they could justify their choices and share their opinion about the impact of code annotations on code readability.

\subsection{Research Questions}

We targeted the following research questions:

\MyPara{RQ1: How does using code annotations affect the perception of code readability?} The goal of this RQ is to investigate whether there is a general preference regarding readability when choosing between a solution based on annotations and an object-oriented alternative. To answer this research question, we evaluated the survey answers, verifying the distribution of responses for each question.

\MyPara{RQ2: Do the usage categories influence developers' perception of the effects of code annotations in readability?} The goal of this RQ is to verify whether the annotation usage categories influence the choice of the participants. To answer this question, we investigated whether participants consistently preferred annotations regardless of the usage category. 

\MyPara{RQ3: Do developers from different demographics perceive the effects of code annotation on readability differently?} The goal of this RQ is to investigate whether the developers' experience affects their perception of annotation's impact on readability. To answer this question, we searched for an association between the answers, the participant's experience (e.g., coding in general, Java, and annotations), and the annotation usage category. We also verified whether participants with a consistent preference for annotations, or their alternatives, had some characteristics in common.
 
\MyPara{RQ4: What factors can influence annotated code readability?} The goal of this RQ is to identify factors that can influence the readability of code with annotations. To answer this question, we performed a qualitative analysis of the answers to the open question wherein the participants spontaneously expressed their opinions about the effects of annotation on code readability. 

%After the evaluation of the categories by the preliminary study, the survey was designed according to the study goals described in this section. The details of the survey design and the reasoning behind the important choices are described in section \ref{sec:survey}. We report the survey results in Section \ref{sec:results} and discuss the implications of this work in Section \ref{sec:discuss}. 

\subsection{Survey Structure}
\label{sec:survey}

The survey starts by asking for the participants' consent, followed by questions about demographic information (9 questions), in which we ask about gender, age, education level, occupancy, programming experience, and knowledge about code annotations. Afterward, the participants answered 15 questions that presented pairs of equivalent code snippets, one without annotations and the other using annotations. Each participant received a random order of the set of 15 questions to avoid biases related to the sequence of readability questions. All the participants received all 15 questions. A final section presents an open question in which the participants could express factors influencing the annotated code's readability and another one where they could provide final comments for us. In total, participants answered 25 questions. The final instrument is available in the replication package.

\subsection{Code Snippets for Readability Comparison}

\begin{figure}[ht]
    \centering
    \includegraphics[width=\linewidth]{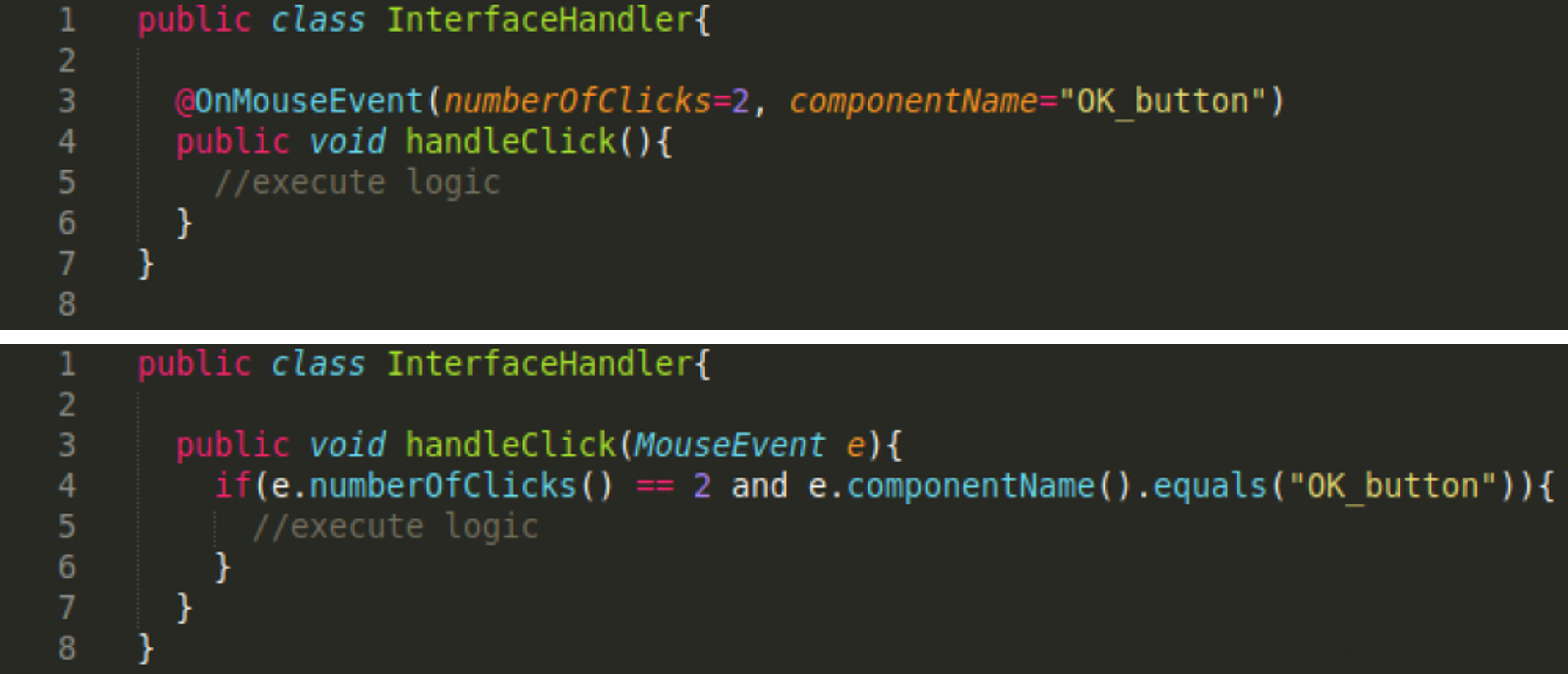}
    \caption{Code snippets of a question of the Callback Method category, with and without annotations}
    \label{fig:callback}
\end{figure}

\begin{figure}[ht]
    \centering
    \includegraphics[width=250pt]{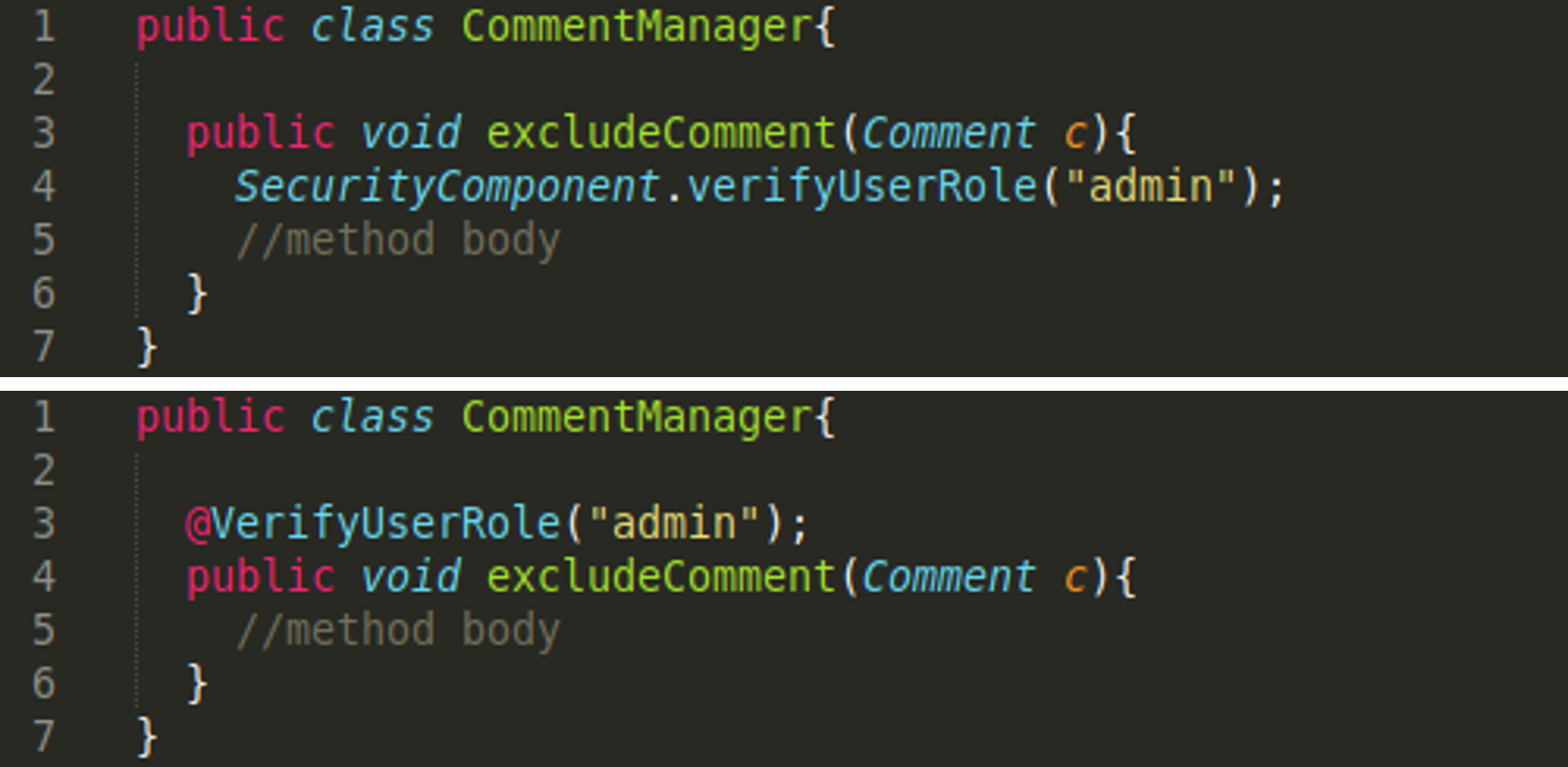}
    \caption{Code snippets of a question of the Proxy Configuration category, with and without annotations}
    \label{fig:proxy}
\end{figure}

To focus the comparison of snippets in the usage of annotations, we designed the snippets to have similar characteristics, especially those known to influence code readability \citep{Buse2010, Posnett2011, tashtoush2013impact}. So, we considered the following guidelines when designing the code snippets:

\MyPara{Small code snippets with similar size:} 
The two code snippets presented to the participant ranged from 4 to 18 lines of code, including blank lines for better code organization. The difference between the number of lines of code from both snippets ranged from 0 to 3. The code snippets omitted parts irrelevant to the comparison, which were substituted by comments containing placeholders, such as \texttt{method body} or \texttt{software logic}, letting participants focus on the differences in the code structure. The comment's placeholders were present on both code snippets (examples in Fig. \ref{fig:callback} and Fig. \ref{fig:proxy}).

\MyPara{Highlighting and formatting:}
Since code highlighting \citep{beelders2016syntax} and formatting \citep{codehard} can influence the perception of readability, we used the same approach in all code snippets. 

\MyPara{Similar wording:}
The exact words were used in both code snippets to compose the code based on annotations and without them, facilitating the mapping of elements between code snippets and reducing the influence of the API vocabulary \citep{lawrie2006s} in the answers.

\MyPara{Category distribution:}
To cover different scenarios, we considered the five usage categories for code annotations presented in Section \ref{sec:categories} and created three questions for each type. This number of questions aims to provide different scenarios inside each category and diversify domains and applications. As presented in Section \ref{sec:categories}, we identified the usage categories based on documented patterns and evaluated their presence in frameworks and APIs (Section \ref{sec:annotations_usage}), reducing the threat of including categories that do not have a significant occurrence in practice or failing to include a relevant category not documented in the literature.

%in the study , in which we evaluated if existing APIs and frameworks have a significant occurrence of annotations from each category. Since our goal is to include in the survey usage scenarios that represent different usages of annotations in practice, this evaluation study aims to reduce the threat of including a category that does not have a significant occurrence in practice or failing to include a relevant category not documented by the literature. 

\MyPara{Designed to represent real scenarios:}
We designed the code snippets based on domains with real-world metadata-based frameworks. 

The code annotations do not add or change the behavior of a class but add metadata processed externally by a class or component that reads the annotation. Because of that, in all snippets from our survey, the behavior depends on an external library. Since that process is transparent to the users of the annotation API, it would not make sense to include the code that reads and processes the annotation in this comparative analysis. Therefore, since the actual behavior of the code could not be determined just by what is in the snippet, we stated to the participants that both snippets being compared had the same behavior.

To ensure behavior equivalence, when designing the questions, we made sure that: (a) the same information defined through the annotations was available using another approach for the external component in the other snippet; (b) the external logic that processes the annotations was called in the same sequence as the equivalent object-oriented alternative. For this second condition, the respective pattern structure was used as the basis to define the object-oriented equivalent solutions. The alternatives depend on the category since metadata is used for a different purpose for each of them. In the following, we discuss examples for each category. We refer the reader to the replication package for the complete list of code snippets.

\begin{figure}[ht]
    \centering
    \includegraphics[width=250pt]{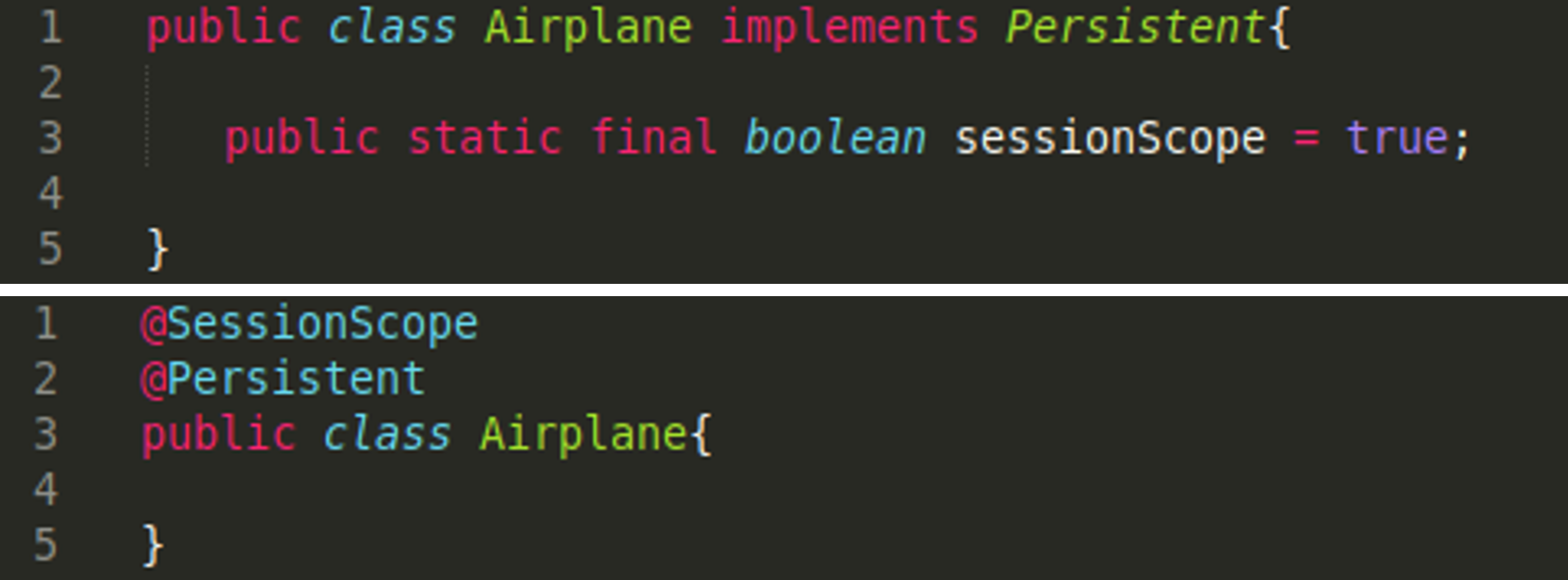}
    \caption{Code snippets of the Rule Definition category, with and without annotations}
    \label{fig:rules}
\end{figure}

The annotations used for \textbf{Rule Definition} in classes can have as alternatives an instance method that returns a fixed value, static fields with default names, and marker interfaces\footnote{\url{https://en.wikipedia.org/wiki/Marker_interface_pattern}} (like \texttt{Serializable}). Fig. \ref{fig:rules} presents a question used in the survey that exemplifies using a maker interface and a static field as an alternative to annotations.
  
For the \textbf{Callback Method} category, which configures a method invoked in response to a situation, the alternative adopted in an object-oriented approach might use a conditional statement inside the method verifying the conditions for handling it. Fig. \ref{fig:callback} presents the code snippets from a question used in the survey in which the metadata in the annotation defines a more fine-grained condition for calling the method. In this case, the same information is used in the conditional statement. 

\begin{figure}[ht]
    \centering
    \includegraphics[width=250pt]{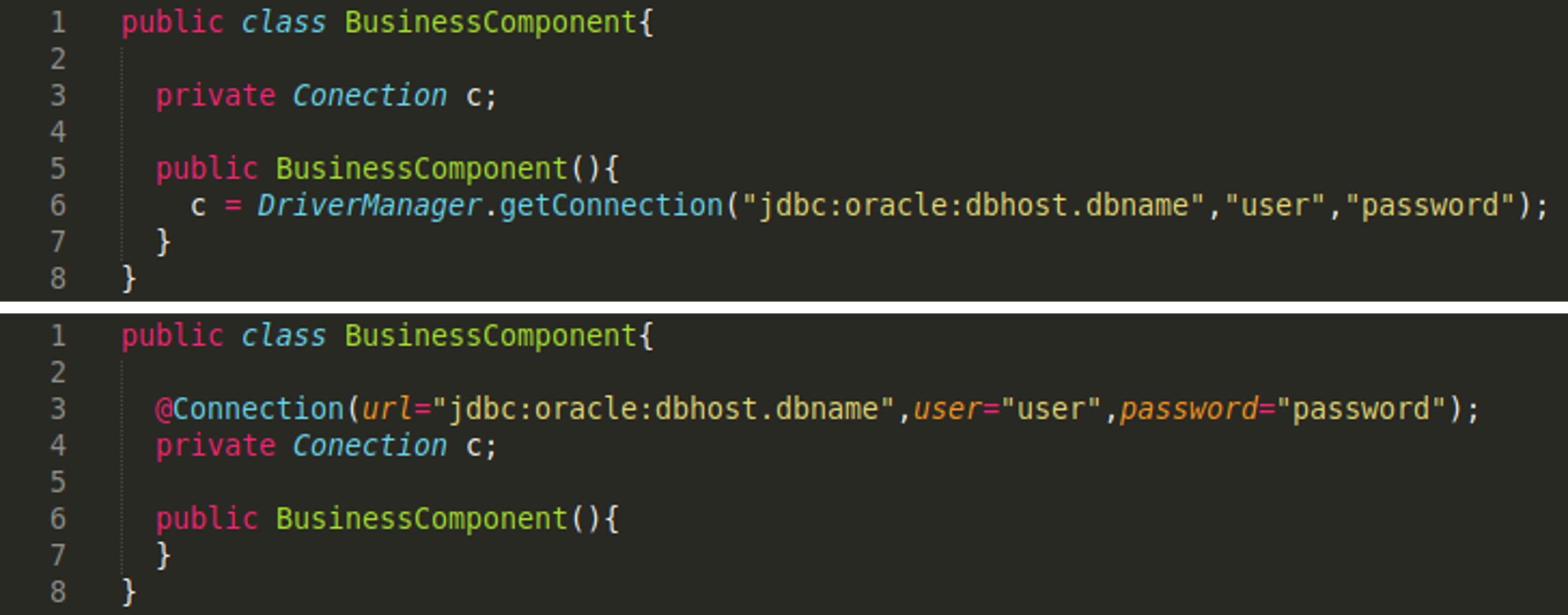}
    \caption{Code snippets of the Dependency Injection category, with and without annotations}
    \label{fig:injection}
\end{figure}

Since \textbf{Proxy Configuration} uses metadata to drive the behavior of dynamic proxies, aspects, or interceptor components, a simple object-oriented alternative is to add a method call with the same parameters at the method's beginning or end. Fig. \ref{fig:proxy} presents a survey question in which an annotation defines an access control constraint to execute the method. For \textbf{Dependency Injection}, since the metadata is used to define what instance should be injected, alternatives might use factories to retrieve the dependency. Fig. \ref{fig:injection} presents both snippets for a scenario in which the object to be injected represents a database connection. 

\begin{figure}[ht]
    \centering
    \includegraphics[width=250pt]{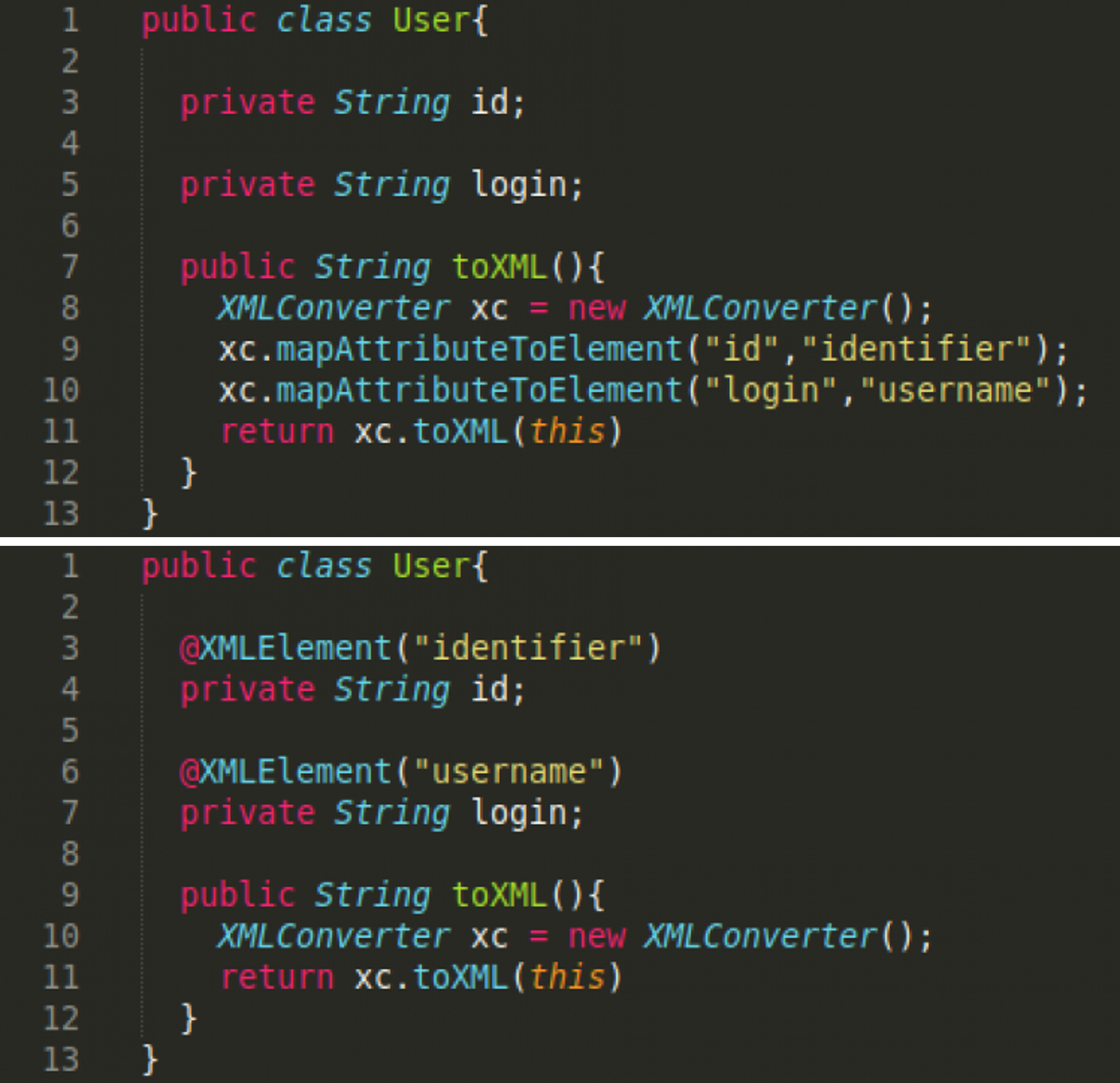}
    \caption{Code snippets of the Information Mapping category, with and without annotations}
    \label{fig:mapping}
\end{figure}

As the last category, alternatives for \textbf{Information Mapping} usually set the information directly in the component that processes the mapping information. In this case, method calls define the mapping instead of the annotations. Fig. \ref{fig:mapping} presents the code snippets from a question used in the survey, in which the annotations are used to define the mapping from a class to an XML format.

Metadata definition in external files is also an alternative, which is frequently used for \textbf{Dependency Injection} and \textbf{Information Mapping}. However, since the scope was to compare with object-oriented alternatives, we excluded using external sources as alternatives for the survey questions.

To ensure that both snippets could lead to the same behavior, we used the same information defined in the annotations in other parts of the code, considering where and how this metadata is used for each category. All the questions were discussed internally among the seven researchers, and an agreement about the equivalence of the code snippets was reached before proceeding to the pilot study. The pilot study participants were asked to provide feedback about the equivalence of the code snippets, and no issue was raised.  

All the images from the code snippets presented in this section (Figs. \ref{fig:rules} - \ref{fig:mapping}) are the same ones used in the survey. The reader can notice the presence of the same code highlighting and line numbering in both options.

\subsection{Answers for the Readability Questions}

For each question, the participants chose one of the following options:

\begin{itemize}

    \item CODE A is definitely more readable.
    \item CODE A is slightly more readable.
    \item Similar readability, but I prefer CODE A.
    \item Similar readability, but I prefer CODE B.
    \item CODE B is slightly more readable.
    \item CODE B is definitely more readable.
    
\end{itemize}

We choose not to include a neutral answer, forcing the participants to choose a side, even if stating that the readability is similar. This approach aims to emulate practice, where a developer must decide how to design a piece of code, even without a clear preference. In the analysis, we considered both answers in the middle (Similar Readability but preferred with/without annotation) when referring to neutral answers. For each participant, we randomized the order of the questions and whether the annotated code would be shown on the left (Code A) or the right (Code B). 

Since the answers represent an ordinal scale, to consider the intensity of the readability impact and to calculate an average score, numeric values were associated with them \citep{briand1996application} as follows: -5; -3; -1; +1; +3; +5. We distributed the values by the same interval, using negative values for answers that considered the code without annotations more readable. This score can be summed and divided by the number of answers to get the average score for a given question or participant and gives an approximation of the overall preference in the question.

\subsection{Survey Analysis}

To address \textbf{RQ1} (overall preferences), we analyzed the distribution of answers for every question. For \textbf{RQ2} (influence of scenarios), we used the average score for each participant. To answer \textbf{RQ3} (influence of demographics), we compared the score distribution among the questions using Kruskall-Wallis-tests and post-hoc Dunn tests with Benjamini-Hochberg for p-value correction \citep{benjamini1995controlling}. We also used the Chi-square test to verify an association between participant demographics (e.g., age, gender, programming experience, annotation experience, coding experience, and occupation) and annotation preferences (five categories created based on average participants' scores). We calculated the average scores for every question and participant. We divided the participants into five categories based on their average score: hate($< -3$), dislike ($ -1>$ and $\geq -3$), neutral ($\leq 1$ and $\geq -1$), like ($\leq 3$ and $>1$), and love ($>3$).

To address \textbf{RQ4} (factors), we analyzed the answers to the open question about how annotations affect the code's readability. From the answers, we identified (1) attributes on readability; (2) comments on the specific cases related to the usage categories as presented in Section \ref{sec:categories}, and annotation strategies recommended by respondents. We started with 57 attributes on readability, 92 specific cases, and 67 strategies. 

We categorized the responses using a card sorting approach \citep{Spencer2009}. While grouping the attributes, specific cases, and strategies, we started by reading each one and grouping those representing the same code. We then categorized them into higher-level clusters based on similarities between the meanings of the codes. For instance, when coding the attributes on readability, participants mentioned ``abstraction,'' ``encapsulation,'' ``modularity,'' ``cohesion,'' ``organization,'' etc. These mentions were grouped into  \textit{Design}. When participants mentioned a ``learning curve'' to use annotations, ``fast',' ``comfortable,'' ``intuitiveness,'' ``making the code more understandable,'' and ``ease to read,'' we used the grouping category \textit{Info Processing}.

The whole process of coding and grouping was conducted using continuous comparison \citep{continuos_comparison} and discussion until reaching a consensus. Two researchers jointly analyzed each answer and applied codes. Finally, a third researcher inspected the classification. Our findings aim to complement results from the literature \citep{guerra2013qualitative, guerra2020}, gaining an understanding of the rationale for preferring or not preferring annotations. 

\subsection{Pilot Studies}

Before the widespread distribution of our questionnaire, we conducted two pilot rounds to identify eventual problems and how well participants understood the code snippets. First, two authors of this paper who did not participate in the survey design answered the questionnaire and suggested improvements to the instructions to the participants. In the second round, $28$ people from $6$ countries and $2$ continents answered the questionnaire: experienced developers ($10$), Ph.D. in Software Engineering ($4$), graduate students in Software Engineering ($12$), and undergraduate students ($2$) in Computer Science. Their answers and feedback helped us improve the presentation and clarify some questions. For instance, a participant suggested changing \texttt{sessionVariable} and \texttt{@SessionVariable} to \texttt{sessionScope} and \texttt{@SessionScope}, respectively, in a specific code snippet. Moreover, we asked them to report the time to complete the questionnaire. Based on the average time from the pilots, we informed participants of an estimated completion time in the survey instructions.

\subsection{Recruitment}
\label{survey_recruitment-anlysis}

Our target population was not restricted to developers experienced in annotations or a specific programming language since we wanted to investigate perceptions of different profiles of developers.

First, we advertised the survey on social media sites (including Twitter, Facebook, and LinkedIn) ($19\%$ of the survey responses). Our posts were re-shared multiple times by other participants. Second, we advertised the survey to mailing lists from six different universities ($37\%$ of the responses), as well as mailing lists from Java developer communities ($27\%$) and mailing lists from researchers in computer science ($8\%$). Finally, we advertised it in development and research groups in messaging apps ($9\%$). We avoided emailing a participant directly and scraping email addresses from software repositories since this practice is perceived as ``worse than spam'' \citep{Sebastian:ESEM:2016}.

We obtained $499$ answers. The survey was available between March 24 and July 11, 2021. From the total of $499$ responses, we removed $98$ in which the participants did not answer more than one question ($6\%$ of the survey) and another $69$ responses wherein they filled the survey in a few seconds. Thus, in our analysis, we considered $332$ valid answers.

From these valid answers, we had a balanced number of professional developers (151) compared to students and researchers (166). Most respondents are between 18 and 44 years old (298). Our respondents self-described themselves with advanced/expert knowledge in Java (143) and some familiarity with annotation (287). The gender identification of respondents is as follows: $294$ identified as men, $34$ as women, $7$ as non-binary/gender diverse, and $3$ preferred not to answer.

\subsection{Replication Package}

A comprehensive \textbf{replication package} including our anonymized dataset, scripts, coding process, and the questionnaire is available in the Zenodo\footnote{\url{https://doi.org/10.5281/zenodo.5396378}} open data archive.

\section{Results}
\label{sec:results}

\subsection{RQ1: How does the usage of code annotations affect the perception of code readability?} 

\begin{figure*}[ht]
    \centering
    \includegraphics[width=\textwidth]{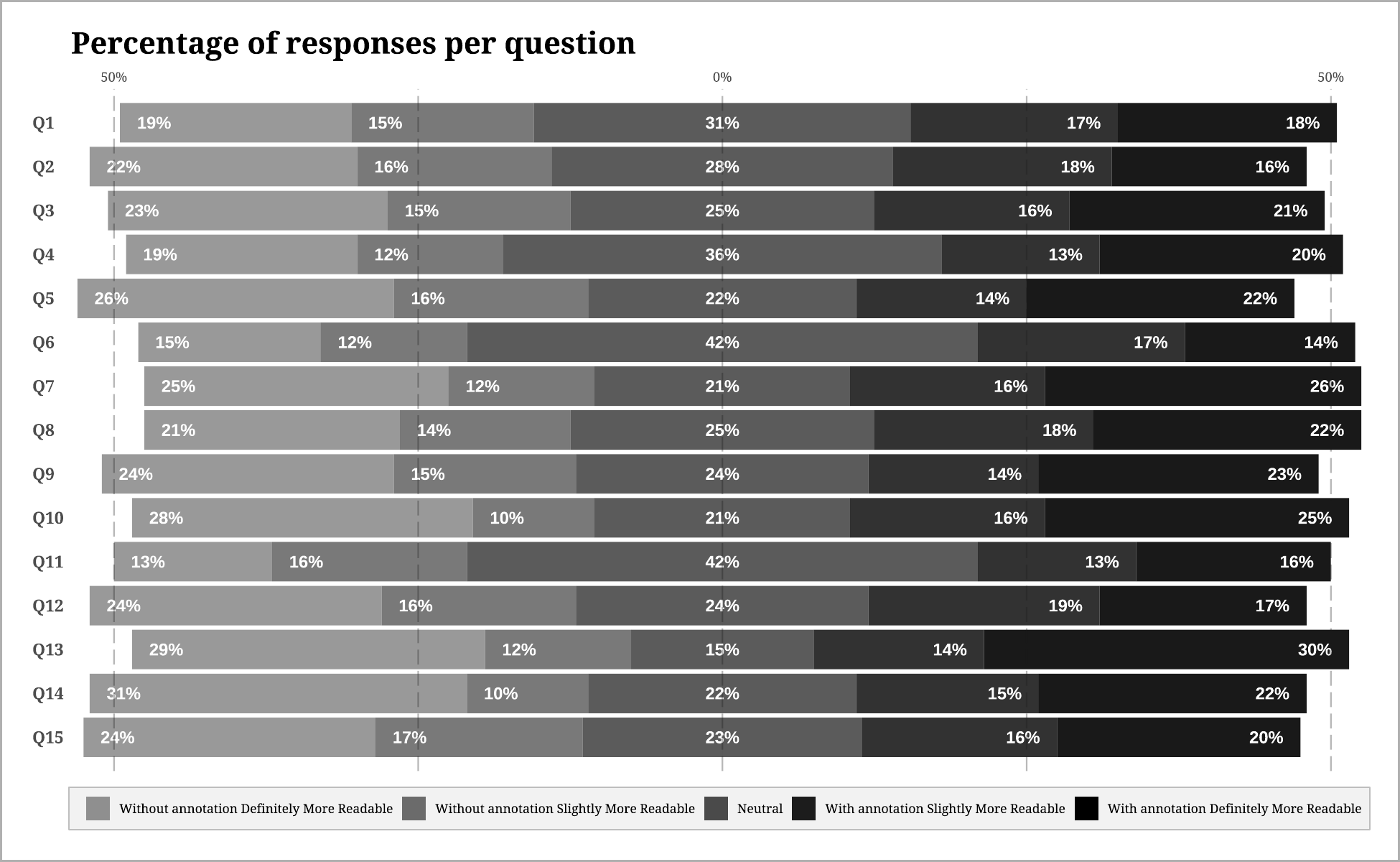}
    \caption{Likert-scale percentage of responses for survey questions}
    \label{fig:likert}
\end{figure*}

To answer this question, we analyzed the total percentage of responses for every question, as presented in Figure \ref{fig:likert}. We can observe that in most questions the total percentage of participants that answered neutral (similar readability but prefer one case or the other) are between 20\% and 30\%, and for questions $Q6$ and $Q11$, this percentage was 42\%. By comparing the non-neutral questions, we can observe that the total number of participants who answered positively about annotations is very similar to those who had a negative perception. The most significant difference was in question Q8, where 56\% of the answers were positive to annotations and 44\% negative. Most responses show differences of 1\% to 2\% in the total number of answers that prefer annotations \textit{versus} prefer without annotations. 

\begin{figure}[ht]
    \centering
    \includegraphics[width=\textwidth]{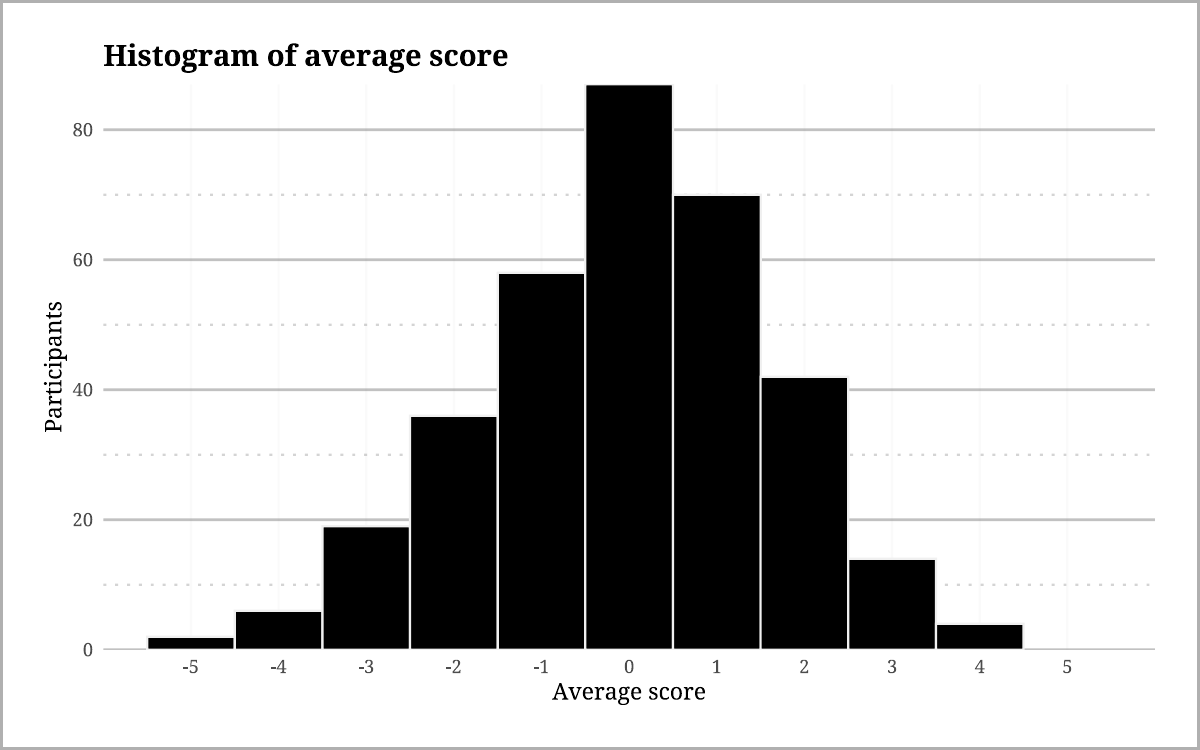}
    \caption{Number of participants per average score range}
    \label{fig:histogram}
\end{figure}

Figure \ref{fig:histogram} displays the average score per question \textit{vs.} the number of participants that had a score between intervals, e.g., we had two participants with a score between $-5$ and $-4.47$. We can see that the result is similar to a normal distribution (Shapiro-Wilk normality test, p-value = 0.4171). By analyzing Figure \ref{fig:histogram}, we could conclude that a significant group of participants revealed a consistent personal preference in their answers.

Considering the ``Neutral'' participants, we identified the ones with two particular kinds of behavior, considering the average value of the absolute score for the answers. A high number represents a participant who generally strongly prefers one of the sides, even being considered neutral on average. In this group labeled ``Neutral with a strong opinion,'' we have 54 participants with an average absolute score between 3.5 and 5. In the other group labeled ``True neutral'', we included the ones with the average value of the absolute score between 0 and 1, which are the ones that assessed the presence of annotations indifferent in terms of readability for most of the questions. Only 33 participants (10\% of the sample) were classified in this group. This shows that even with many participants being ``Neutral'' on average, most chose one of the sides for most of the questions.

We could not reach a consensus about some survey participants' preferences since the statistical tests did not find any differences in the distribution of the answers (Kruskall-Wallis, p-value = 0.5092). In other words, participant preferences were not consistent regarding the usage of code annotations.

\subsection{RQ2: Do the usage categories influence developers' perception of the effects of code annotations on readability?}

To answer this question, we calculated the average score of each participant, grouping the three questions asked for each category. We also classified each person into five preferences categories according to their average score: \textbf{Love} (annotation) with average score $>$ to +3, \textbf{Like} between $>$ 1 and $<=$ 3, \textbf{Neutral} between $>=$ 1 and $<=$ -1, \textbf{Dislike} $>$ -1 and $<=$ -3 and \textbf{Hate} (annotation) $>$ to -3.

Table \ref{tab:categories} presents the number of participants grouped by preferences for each one of the annotation categories asked in the survey. To examine the relationship between participants' preferences and the Annotation Category, we used a Chi-square independence test after verifying its assumptions (the sample consisted of independent observations, and the count in each cell was larger than 5). A p-value less than 0.05 (p-value $=$ 2.2e-16) indicates statistical significance, which means that preferences are associated with the annotation category.

We can notice less neutral preferences for four categories: Dependency Injection, Information Mapping, Callback Method, and Rule Definition. In these four categories, even having a balanced distribution between positive and negative, the higher number of answers are in the extremes. Moreover, participants showed neutrality when using annotation in the Proxy Configuration scenario. The distribution is different in this case, and many answers are neutral.

For the Proxy Configuration scenario, the annotation parameterizes a behavior that can happen before and/or after the target method execution. In the proposed questions, the object-oriented alternative invoked a method with the desired behavior at the target method's beginning and/or end. The participants compared one snippet that defined the parameters declaratively above the target method with an annotation, with a method invocation passing the same parameters inside the target method body. Since we used the same nomenclature, the difference between the snippets might seem to be just a similar line of code defined in different places. That can be a possible reason for this neutrality.  

We also grouped the questions by category: \textit{Proxy Configuration} (Q1, Q2, Q3); \textit{Dependency Injection} (Q4, Q5, Q6); \textit{Information Mapping} (Q7, Q8, Q9); \textit{Callback Method} (Q10, Q11, Q12); and \textit{Rule Definition} (Q13, Q14 e Q15). The answers distributed from each group and each question individually were compared to the general behavior. We could not find a statistical difference between any category or individual question that can indicate that they have a positive or negative impact on code readability (Kruskal-Wallis-test, p-value = 0.5092). This fact can also be observed visually in the chart from Figure \ref{fig:likert}.

\begin{table}[ht]
\caption{Participants preferences by Annotation Category}
    \adjustbox{max width=\textwidth}{
    \centering
\begin{tabular}{l|r|r|r|r|r|}
\hline
                 & \multicolumn{1}{l|}{\textbf{Proxy Configuration}} & \multicolumn{1}{l|}{\textbf{Dep. Injection}} & \multicolumn{1}{|l}\textbf{{Information Mapping}} & \multicolumn{1}{|l} \textbf{{Callback Method}} & \multicolumn{1}{|l} \textbf{{Rule Definition}} \\ \hline
Love            & 22                                      & 96                                      & 106                                     & 92                                  & 110                                 \\\hline
Like    & 67                                      & 33                                      & 35                                      & 35                                  & 28                                  \\\hline
Neutral          & 137                                     & 63                                      & 58                                      & 61                                  & 36                                  \\\hline
Dislike & 75                                      & 32                                      & 23                                      & 44                                  & 25                                  \\\hline
Hate            & 26                                      & 103                                     & 105                                     & 95                                  & 128  \\ \hline                              
\end{tabular}
 }
    \label{tab:categories}
\end{table}

   \begin{table}[ht]
    \caption{Participants demographics grouped by average score}
    \adjustbox{max width=\textwidth}{
    \centering
    \begin{tabular}{|l|r|r|r|r|r|}
        \hline
        \rowcolor{Gray}
        Java Experience & Hate & Dislike & Neutral & Like & Love \\ \hline
        Advanced        & 5	    & 28	            & 44	     & 22	            & 3     \\ 
        Expert          & 1	    & 12	            & 19	     & 6	             & 3     \\ 
        Intermediate    & 3	    & 26	            & 62      & 29	            & 1     \\ 
        Novice          & 1     & 12             & 41      & 20             & 0     \\
        \hline
        \rowcolor{Gray}
        Annotation Experience & Hate & Dislike & Neutral & Like & Love \\ \hline
        Extremely familiar    & 4     & 17             & 36	     & 15             & 4	    \\ 
        Not familiar          & 1     & 15             & 19	     & 10	            & 0	    \\ 
        Slightly familiar     & 3     & 26	            & 45	     & 26	            & 1	    \\ 
        Somewhat familiar     & 2     & 20             & 66      & 26             & 2     \\
        \hline
         \rowcolor{Gray}
        Coding Experience & Hate & Dislike & Neutral & Like & Love \\ \hline
        1 to 2 years                    & 1     & 4              & 9       & 9              & 0     \\ 
        2 to 3 years                    & 0     & 12             & 26      & 11             & 0	    \\ 
        3 to 5 years                    & 0     & 16             & 29      & 11             & 0     \\ 
        5 to 10 years                   & 1     & 19             & 36      & 15             & 2     \\ 
        10 to 20 years                 & 4     & 16             & 41      & 24             & 4     \\ 
        More than 20 years                       & 4     & 11             & 25      & 7              & 1     \\
        \hline
        \rowcolor{Gray}
        Gender  & Hate & Dislike & Neutral & Like & Love \\ \hline
        Man                    & 9     & 66              & 140      & 72              & 7     \\ 
        Woman                    & 0     & 10             & 20      & 4             & 0	    \\ 
        Non-binary                    & 1     & 1             & 5      & 0             & 0     \\ 
        Prefer not to say                   & 0     & 1             & 1      & 1             & 0     \\ 
        \hline
        \rowcolor{Gray}
                Occupation  & Hate & Dislike & Neutral & Like & Love \\ \hline
        Developer         & 6     & 33              & 78      & 29              & 5     \\ 
        	Student	       & 1     & 35             & 56      & 28             & 0	    \\ 
        Researcher	          & 2     & 8             & 19      & 16             & 1     \\ 
  %      Other                   & 1     & 2             & 13      & 4             & 1     \\ 
        \hline
        \rowcolor{Gray}
         Age  & Hate & Dislike & Neutral & Like & Love \\ \hline
 %       Prefer not to say         & 0     & 0              & 0      & 1              & 0     \\
        18 to 24	       & 1     & 30             & 48      & 29             & 1	    \\ 
        25 to 34	          & 5     & 23             & 59      & 18             & 1     \\ 
        35 to 44                   & 3     & 19             & 40      & 23             & 4     \\ 
        45 to 54	          & 1     & 3             & 12      & 6             & 1     \\ 
        55 to 64                   & 0     & 2             & 6      & 0             & 0     \\
        Over 64                   & 0     & 1             & 1      & 0             & 0     \\
        \hline
%         Education Level  & Hater & Dislike & Neutral & Like & Lover \\ \hline
 %       Ph.D.	  & 1     & 11              & 18      & 16              & 1     \\
%        Master's degree		    & 5     & 15             & 40      & 19            & 1	    \\ 
%        Undergraduate degree    & 4     & 28             & 70      & 20             & 5     \\ 
%        High school	               & 0     & 22             & 31      & 16             & 0     \\ 
 %       Technical training	 	  & 0     & 1             & 5      & 6             & 0     \\ 
  %      No formal education    & 0     & 1             & 2      & 0             & 0     \\
 %       \hline
    \end{tabular}
    }
    \label{tab:demographics}
\end{table}

\subsection{RQ3: Do developers from different demographics perceive the effects of code annotation on readability differently?}

To answer this question, we considered the data presented in Table \ref{tab:demographics}. We grouped the participants from each average score category by Java experience, annotations experience, coding experience, gender, occupation, and age. By performing the chi-square test, all the p-values returned are larger than the significant level (p-value $>$ 0.05). Thus, there is insufficient evidence to conclude that any demographic is associated with annotation preferences. 

\subsection{RQ4: What factors can influence annotated code readability?}

Participants suggested several factors that make annotations positively or negatively impact code readability, as summarized in Table \ref{tab:attributes-cases} and Table \ref{tab:attributes-cases}. The most common explanations were related to effects in code characteristics and quality attributes (94+/20-)\footnote{This notation denotes the number of mentions in which the impact was classified as positive or negative.}. The most common reference was objectiveness (40+/2-), as illustrated by P57: ``Helpful when it removes unnecessarily verbose''. Also related to objectiveness, less duplication (10+) was frequently mentioned: ``usage of annotations can often make the code more readable by reducing the number of repetitive constructions and avoiding duplication code'' (P139). 

Clean code or clarity (23+/10-) was the most controversial topic, with 23 participants mentioning a positive impact while 10 mentioned negative impacts. As P169 stated: ``[annotations] makes it less clear what is going on''. Participants generally explained that the lack of clarity was related to making business logic hidden, what they called ``magic.'' For instance, P236 mentioned: ``I am not so used to using annotations, so they often leave me with the feeling that some `magic' is going on''. Other explanations for the lack of clarity were connected to the overuse of annotations and the choice of poor names. Still related to clarity, participants also mentioned positive and negative effects on explicitness (13+/8-). As illustrated by P236, ``being explicit about what is going on makes the code more readable for the inexperienced. For someone who has experience with a system and is familiar with the typically used annotations, they may be helpful.'' However, RQ3's results contradict this statement. 

Impact on design quality attributes (41+/6-) was also frequently mentioned by the participants, mostly positively. In particular, participants mentioned effects on simplicity (17+/1-), modularity (7+), cohesion (5+), organization (3+), abstraction (2+/1-), elegance (3+), encapsulation (1+), flexibility (1+), and testability (-4). The last was the only one in which participants mentioned only negative effects, as illustrated by P392: ``makes code harder to test ... and given that in my experience testability is king, that's a lot of reasons right there to avoid annotations.''

Participants also mentioned that annotations affect readability by making information (33+/25-) more understandable (21+/21-), faster to read (9+), and easier to learn (2+). One can notice the controversy around understandability, with 21 mentions of positive impacts and 21 negative ones. On the negative side, 11 participants discouraged representing logic in annotations, which correlates with the previous findings related to ``magic'' and explicitness. Other attributes mentioned less frequently were maintainability (7+/2-) and productivity (4+). The complete coding book and subcategory counting are available in our replication package, mentioned in Section~\ref{survey_recruitment-anlysis}.

\begin{table}[ht]
\caption{Impact attributes on readability when annotations are used and specific cases reported when using/avoiding annotation}
\centering
\begin{tabular}{|l|r|r|}
        \hline
        \rowcolor{Gray}
        Attributes & Positive(+) & Negative(-)  \\ \hline
        Code & 94 & 20  \\  
        Design & 41 & 06  \\ 
        Info processing & 33  & 25  \\ 
        Logic & 01  & 11  \\ 
        Maintainability & 07 & 02 \\ 
        Productivity & 04 & 00  \\
        \hline
        \rowcolor{Gray}
        Annotation Usage Categories & Positive(+) & Negative(-)  \\ \hline
        Proxy configuration & 17  & 10 \\ 
        Rule Definition & 23  & 03 \\
        Information Mapping & 04 & 00  \\
        Callback Method & 02 & 03  \\  
        Dependency Injection & 02 & 03 \\ 
        \hline
    \end{tabular}
    \label{tab:attributes-cases}
\end{table}

Participants also explicitly mentioned scenarios that could be mapped to the annotation usage categories described in Table \ref{tab:attributes-cases}. Using annotations for defining rules (12+), which receive specific mentions of validation rules, and for information mapping (4+) received only positive recommendations. We could not observe the same results in the other research questions when analyzing each annotation category in isolation. Some of them had mixed recommendations. More specifically, cross-cutting concerns related to the \textit{Proxy Configuration} category received significant mentions both recommending to use and avoid annotations (17+/10-). \textit{Callback Method} (2+/3-), configuration (1+/3-), and dependency injection (2+/3-) also received mixed recommendations.

\begin{table}[ht]
\caption{Annotation usage strategies recommended by respondents}
\centering
\begin{tabular}{|l|r|}
        \hline
        \rowcolor{Gray}
        Strategies & \# of mentions  \\ \hline
        Avoid ``magic'' & 15 / 25.9\%  \\
        Avoid overuse & 09 / 15.5\%  \\
        Keep it simple & 09 / 15.5\%  \\
        Choose good names & 07 / 12.1\%  \\
        Avoid custom annotation & 04 / 6.9\%  \\
        Do not use in simple cases & 03 / 5.2\%  \\  
        Keep it close to the code being annotated & 03 / 5.2\%  \\
        Use few parameters & 03 / 5.2\% \\ 
        Do not mix annotation types & 02 / 3.4\% \\ 
        Be explicit & 02 / 3.4\% \\ 
        Use for average developers & 01 / 1.7\% \\ 
        \hline
    \end{tabular}
    \label{tab:strategies}
\end{table}

Finally, the qualitative analysis also revealed recommendations for using annotations, represented in Table \ref{tab:strategies}. The most mentioned strategy suggests avoiding cases where the behavior driven by the annotation seems obscure or magical to the user (15). In the words of participant P118: ``Annotations are like any abstraction: done well, it improves the readability; when it hides too much or relocates information to the wrong place, it is confusion and decreases readability.'' Participant P125 summarized this idea as ``it helps when the annotation’s behavior is obvious from the way it's used; it hinders when behavior is magic.'' 

Participants frequently mention other strategies, such as keeping annotation usage simple (9) and choosing intuitive names (7). P179 illustrates this idea: ``If the functionality is well encapsulated, and the name is intuitive enough, annotations can improve a lot the readability and make the code much more maintainable, even to someone without experience in that piece of code.'' However, eight additional participants highlight that annotations can be helpful, but overuse should be avoided. As mentioned by participant P179: ``if used everywhere, creating some sort of annotation hell, you think you understand the purpose of that code, but has no clue on how to use it.''

Finally, one may argue that some identified categories do not directly relate to readability. However, we also recognize that readability is subjective, and the participants could consider other maintainability aspects. This qualitative analysis adds to the current literature (e.g., \citep{guerra2013qualitative, guerra2020}) by bringing additional information from spontaneous manifestations of the participants about the readability and design of annotated code. 

\section{Discussion}
\label{sec:discuss}

Based on the survey results, we could not find an association between the usage of annotations and a consistent positive or negative impact on code readability. Searching for correlations between the answers and participant demographics (e.g., programming experience, gender), we also could not find a particular profile with a different tendency. We evaluated the questions grouped by the annotation usage category, including individually, and did not find a statistically different distribution. 

\textbf{A practical implication (i) of these results for API designers is that annotations will not necessarily harm the readability}, as frequently stated in the gray literature \citep{StackOverflow-AgainstAnnotations,Bugayenko2016,Warski2017}. This scenario is also the case if the API has a specific target audience, such as beginners or experts, or has a particular annotation usage, such as \textit{Information Mapping} or \textit{Proxy configuration}.

Our study also found that many participants consistently prefer code annotations (or not). The participants with positive and negative tendencies were balanced. The difference in the number of participants with a strong preference, classified as ``Love'' and ``Hate,'' was not significant. Still, participants classified as ``Like'' and ``Dislike'' comprise around half of the sample, representing a moderated tendency from one side. That finding highlights the importance of the study results as \textbf{a practical implication (ii) to help API designers avoid generalizing their personal views regarding the impact of annotations in code readability to make a decision}.

The two code snippets of each question in the study use the same terms and have a similar size, representing a regular usage of annotations. However, in real projects, annotations can be used in different conditions, and problems in readability might arise not simply from this programming language feature but from how it is used. So, even if our results provide evidence that code annotations do not negatively impact code readability, some practices can improve and reduce the readability of annotated code.

The qualitative study found statements that claimed annotations' positive and negative impact on the code and information processing. For instance, P117 stated that ``some annotations are very clear in their function and improve readability by making code more concise. Others are difficult to understand and serve only to confuse me.'' P139 claimed that ``annotations could often make the code more readable by reducing repetitive constructions and avoiding duplication code. However, over-usage of annotations might become an issue.'' Similarly, P137 said that ``sometimes annotations allow writing less code and keeping related code together .... it requires more time to get used to them because the code is not explicit, but they are faster to read after you know them. But mixing annotations from too many different contexts in the same code seems bad to me.'' That confirms the results from the survey that there are participants with preferences on both sides. Importantly, \textbf{(iii), the participants recommend avoiding abuses in the usage of annotations}.

The abuse of code annotations can be detected using a metric suite that measures characteristics in their usage \citep{lima2018jss}. These abuses can be prevented from the point of view of the API designer and API user. The API designer might use practices that allow a more general definition of metadata, such as General Configuration and Annotation Mapping \citep{guerra2010idioms}, or provide alternatives based on code conventions. The API users can use the alternatives provided by the API to reduce the number of annotations, such as the practices previously mentioned, and avoid concentrating annotations from several frameworks in the same class. 

The impact on design had a high number of mentions (usually on the positive side), which confirms results from previous works that claim a reduction in coupling and the support for a more consistent evolution \citep{guerra2020}. For instance, P244 stated, ``I generally found it to improve readability, most commonly through (a) better modularization and (b) bringing information closer to where it is needed.'' A high number of participants claimed a positive impact on code quality \citep{Yu2019}. For example, P95 claimed, ``Annotations improve code's readability by making clearer and more explicit logic representations.'' This evidence brings us to another practical recommendation \textbf{(iv) that annotation-based API can provide a more consistent code evolution with better coupling and complexity metrics}.

%As for the proposed annotations categories, we observed that they are present and used in real-world projects. We also found that some annotation schemas (a group of related annotations) usually fit into one whole category. Nevertheless, there was some overlapping, and we intend to investigate these occurrences of annotations classified in more than one category, identify specific usages, and define subdivisions.

%Further detailing the overlapping, one example is between \textit{Information Mapping} and \textit{Callback Method} when the mapping occurs between methods and services from other API. The uncertainty is that the method is called due to an event (\textit{Callback Method}) but mapped from a request received from another API. In other instances, the \textit{Information Mapping} also seems to overlap \textit{Dependency Injection}. That is especially true when what is being mapped is also injected into the target class. 

\textit{Proxy Configuration} and \textit{Rule Definition} are the categories most mentioned in our qualitative analysis. \textit{Proxy Configuration} appeared in the qualitative analysis as the usage of annotations for crosscutting concerns, receiving mentions that encourage and discourage using annotations. For instance, P382 stated, ``I think annotations contribute to better readability when considering Dependency Injection or when we plan to decorate a class/method with a feature that can be regarded as crosscutting and not specific to the business rule.'' The \textit{Rule Definition} category appeared only encouraging its usage, which contradicted the general behavior in the questions of this category in the survey. P233 said, ``It positively affects simple aspects such as HTTP requests or validation .... For example, on transaction issues, it is sometimes necessary to perform other activities within a transaction. Using annotations on these parameters helps the readability and clarifies what activities are being carried within this transaction.'' 

Moreover, \textit{Information Mapping} is one of the most familiar categories to developers and practitioners, given well-established frameworks such as JPA (to perform ORM) and JAXB (to serialize/deserialize XML). As stated by participant P46: ``In general, annotations can improve the code's readability, especially in mapping situations.'' Furthermore, P148 said: ``Overall, annotations usage improves code readability when abstracting recurrent tasks like mapping values, validating data or fetching session variables, mostly by substituting statements unrelated to the business logic.'' Some participants, like P143, believe that annotations should be primarily used to map data: ``In my opinion, annotations need to either hide ugly implementation, repetitive details or be related to field technicalities, for instance, use in mapping. However, you do not want to have to think about what they do or go back and forth.''

\section{Limitations}\label{sec:limit}

\MyPara{Sampling bias (survey).} We combined multiple strategies to reach a diverse sample. As described in Section~\ref{survey_recruitment-anlysis}, we achieved a diverse population, especially regarding participant experience. Most participants were from America, despite reaching participants from many countries. We also have a low number of women and non-binary respondents, which are under-represented in software development \citep{gender_development,prana2021including,trinkenreich2022women}. Our results are only valid for our respondents, and we encourage this work to be replicated in different scenarios.

\MyPara{Target population.} In this study, we focused on the perspective of outsiders---participants who were looking at the code for the first time. Even though this perspective is relevant because projects often receive new developers, especially some open source projects \citep{steinmacher2015social}, the results may change when considering the perspective of frequent contributors to the code. Future work can investigate the impact of annotations in this context.

\MyPara{Response biases.} The code snippets were designed based on annotation usage categories whose presence we verified in well-known Java projects, as described in Section \ref{sec:categories}. Nevertheless, we acknowledge that we may not have included all possible usage scenarios and representative samples in the survey instrument. Nevertheless, to mitigate response ordering bias, we randomized the order of the questions and the order of the code snippets. We also used small code snippets with the same vocabulary, similarly highlighting and formatting the code. We also understand that comparing two snippets may reduce potential threats. This strategy was employed in other studies such as \cite{dos2018impacts} and \cite{lucas2019does}. By analyzing the comments provided by participants, we also found cases suggesting that developers carefully inspected both snippets, trying to spot differences between them. For instance, P153 mentioned ``I actually enjoyed taking it as it was interesting to see the differences between implementations with and without annotations.''

\MyPara{Self-selection bias.} Our survey used English as a primary language, which may have influenced the willingness of non-English speakers to participate. Future studies might translate our survey and investigate regional differences.

\MyPara{Confirmation bias.} Confirmation bias is the tendency to interpret, favor, and recall information in a way that supports one's prior beliefs. This bias may have influenced our respondent's answers. To mitigate this bias while recruiting participants and writing the instrument, we characterized the study as a code legibility research, avoiding direct mentions to code annotation.

\MyPara{Sampling bias (Annotation Categories).}  In the annotation usage category evaluation, we included all the annotation schemas from Java standard APIs found and the same number of third-party libraries. However, from the 326 schemas found, only 21 were accepted. While this sample provided enough evidence to evaluate the categories for the survey study, the third-party libraries were misrepresented. Further studies that target evaluating the frequency of annotations in each category should consider a different sampling strategy.

\MyPara{Annotation Categories Evaluation.} We used different researchers to address potential issues such as confirmation and researcher bias associated with the annotation category evaluation. One of the researchers that performed the classification was previously involved in identifying the patterns used as the basis for the categories. After receiving an initial explanation of the categories, a second researcher relied on the categories' descriptions and the patterns used for their definition. To minimize the impact of researcher bias, both researchers independently conducted the classification, and any discrepancies in the initial classification were used to calculate agreement. Subsequently, a consensus was reached through discussion, incorporating the primary sources of disagreement. Furthermore, the author responsible for selecting the projects and annotation schemas did not partake in the classification process.

\MyPara{Qualitative analysis.} We employed qualitative procedures to classify the open-question answers provided by our respondents. These procedures are subject to interpretation bias. Multiple researchers used a constant comparison to mitigate this threat and negotiated an agreement to conduct the analysis. All the researchers involved have experience in qualitative methods and annotation usage in practice.

\MyPara{External validity (Annotation Categories).} A possible threat to survey validity is the misrepresentation of different usages of code annotations. If the survey includes questions representing just a subgroup of scenarios in which annotations can be used, the results can be considered valid only for that scope. In the survey design, we considered annotation usage categories extracted from documented patterns from the literature, including three questions for each category. To ensure that these categories represent annotations present in real frameworks and APIs, we performed the evaluation study reported in Section \ref{subsec:categories-study}, finding a significant occurrence for the five categories considered. 

\MyPara{Construct validity (Pair of Snippets).} A possible threat to survey validity is if the pair of code snippets cannot generate an equivalent behavior. We created the pair of snippets based on the structure of each annotation category, making sure that (a) the same information defined through the annotations is available using another approach for the external component in the other snippet, (b) the external logic that processes the annotations is called in the same sequence as the equivalent object-oriented alternative. The authors extensively discussed all questions internally and agreed on the equivalence of the code snippets before conducting the pilot study. To check our intuition that the pair of snippets were equivalent and could provide the same understanding, we also conducted a pilot study as described in Section \ref{sec:survey}. During the pilot study and the survey, we invited the participants to provide feedback regarding any concerns about the code snippets, but no issues were raised.

%\MyPara{Construct validity.} We based our questions on real-world metadata-based frameworks distributed in five proposed annotation usage categories to enhance the construct validity. They were proposed based on documented patterns \citep{guerra2016design} and verified the occurrences of such categories in real-world Java projects, as described in Section \ref{subsec:categories}. We also employed pilot studies to test and collect feedback about our instrument.

\section{Conclusion}\label{sec:conclusion}

This paper reported a study investigating the impact of Java annotations on code readability. We did not identify a positive or negative impact, even when segmenting the analysis for usage scenarios or demographic characteristics. The personal preference in favor or against annotations in around half of the participants reinforces the idea that practitioners may not consider their perceptions as true for other developers when making a decision. 

To conduct the survey and elaborate our readability questions, we considered five annotations usage categories based on existing patterns. We evaluated the categories in a study that classified annotations from real-world schemas and found that all annotations could be classified into one of the categories and that all the categories had a significant occurrence in the annotation schemas. We consider the insights obtained from this study as a secondary result of this work. 
%In the readability study, these categories were used in the design of the questions to cover the usage of annotations in different scenarios. 

While several factors should be evaluated when deciding to use or avoid annotation in the design of an API, our results revealed that readability is a quality attribute for which there is no consensus, and additional care should be taken. It is worth highlighting that participants also provided recommendations on scenarios in which to use annotation (e.g., validation, mapping, and rule definition) and not to use it (e.g., business logic). There were also mentions that developers should avoid overuse of annotation, keep the annotation usage simple, and provide good names. In short, from these results, we could extract four practical implications that annotation-based API developers can use to improve annotation readability.

Future studies can investigate readability and other practices in the design of annotated APIs. For instance, the presence of classes with a high number of annotations, with large annotations \citep{lima2018jss}, and their repetition on similar elements \citep{teixeira2018does}, can be responsible for the unfavorable positioning of some participants about annotations. Considering that annotations alone are not responsible for reducing readability, future work can explore how practices related to annotations usage can affect this property.

\section*{Data Availability}
\label{sec:dataavailability}

All the data related to this research, including an anonymized dataset, scripts, and the questionnaire, is available in the Zenodo (\url{https://doi.org/10.5281/zenodo.5396378}) open data archive.

\section*{Conflict of Interest}
The authors declared that they have no conflict of interest.

\section*{Acknowledgments}
We thank all the respondents who spent their time answering our survey. We expect the results to benefit the developers and API designers and inspire new research on this topic. This work is partially supported by FAPESP (grant \#2019/12743-4), CNPq/MCTI/FNDCT (grant \#408812/2021-4), MCTIC/CGI/FAPESP (grant \#2021/06662-1), and NSF (grants 2236198, 2247929, and 2303042).

\bibliographystyle{spbasic}      % basic style, author-year citations
\bibliography{references.bib}   % name your BibTeX data base

\end{document}